\documentclass[epj-spec]{svjour}
\usepackage{graphics}
\usepackage{graphicx, subfig}
\usepackage{epstopdf}
\usepackage{amssymb}
\usepackage{amssymb,amsmath,amscd}

\let\trueint=\int
\let\truesum=\sum
\def\int{\mathop{\textstyle\trueint}\limits}
\def\sum{\mathop{\textstyle\truesum}\limits}

\let\<=\langle
\let\>=\rangle

\renewcommand\labelitemi{\ifmmode\circ\else$\circ$\fi}
\begin{document}
\title{Mach Reflection and KP Solitons in Shallow Water}
\author{Harry Yeh\inst{1}\fnmsep\thanks{\email{harry@engr.orst.edu}} \and Wenwen Li\inst{1} \and Yuji Kodama\inst{2}\fnmsep\thanks{\email{kodama@math.ohio-state.edu}}}
\institute{School of Civil $\&$ Construction Engineering, Oregon State University, OR 97331 \and Department of Mathematics, Ohio State University, Columbus, OH 43210}

\abstract{
Reflection of an obliquely incident solitary wave onto a vertical wall is studied analytically and experimentally. We use the Kadomtsev-Petviashivili (KP) equation to analyze the evolution and its asymptotic state. Laboratory experiments are performed using the laser induced fluorescent (LIF) technique, and detailed features and amplifications at the wall are measured. We find that proper physical interpretation must be made for the KP solutions when the experimental results are compared with the theory under the assumptions of quasi-two dimensionality and weak nonlinearity.  Due to the lack of physical interpretation of the theory, the numerical results were previously thought not in good agreement with the theory. 
With proper treatment,  the KP theory provides an excellent model to predict the present laboratory results as well as the previous numerical results.  The theory also indicates that the present laboratory apparatus must be too short to achieve the asymptotic state. The laboratory and numerical results suggest that the maximum of the predicted four-fold amplification would be difficult to be realized in the real-fluid environment.
The reality of this amplification remains obscure.
} 
\maketitle
\section{Introduction}
\label{intro}
Due to the nonlinearity effect of surface waves, the interference of multiple waves causes local amplification. In deep water, anomalously large waves are often called ÒfreakÓ (or rogue) waves. The cause of this phenomenon is not fully understood, but has been speculated by various theories. A traditional explanation is the role of currents. When strong currents meet waves moving in the opposite direction, the currents focus and concentrate sets of waves, causing individual peaks that are exceptionally higher than their surroundings. The amplification can also occur when disparate trains of waves meet together. Such intersections sometimes cause waves to be higher and steeper. 

Freak-wave formation in shallow water was analyzed by Pelinovsky, et al. \cite{PTK:00} within the framework of nonlinear-dispersive theory in single propagation space. They attributed the main mechanism of freak-wave formation to the spatial-temporal focusing of radiative wave packets and a solitary wave due to the difference in their propagation speeds. They pointed out that a freak wave is rare, short-lived, and always has a small ratio of nonlinearity to dispersion.

\begin{figure}[t]
\begin{center}
\includegraphics[width=3.2in,height=2.2in]{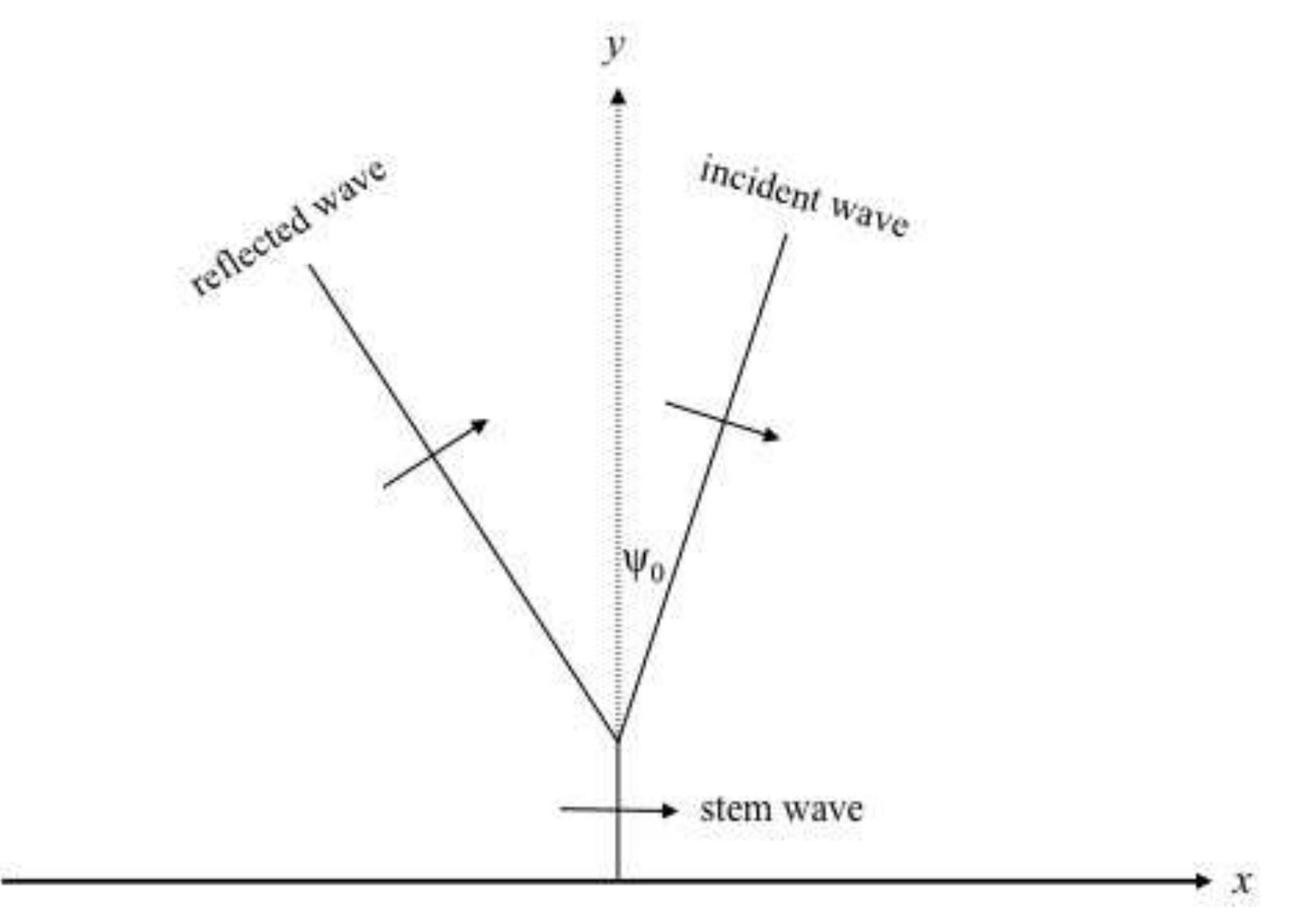}
\end{center}
\caption{Definition sketch for Mach reflection: $\psi_0$, incident wave angle.}
\label{fig:Fig1}
\end{figure}

\begin{figure}[t]
\centering
\includegraphics[width=4.5in,height=2.5in]{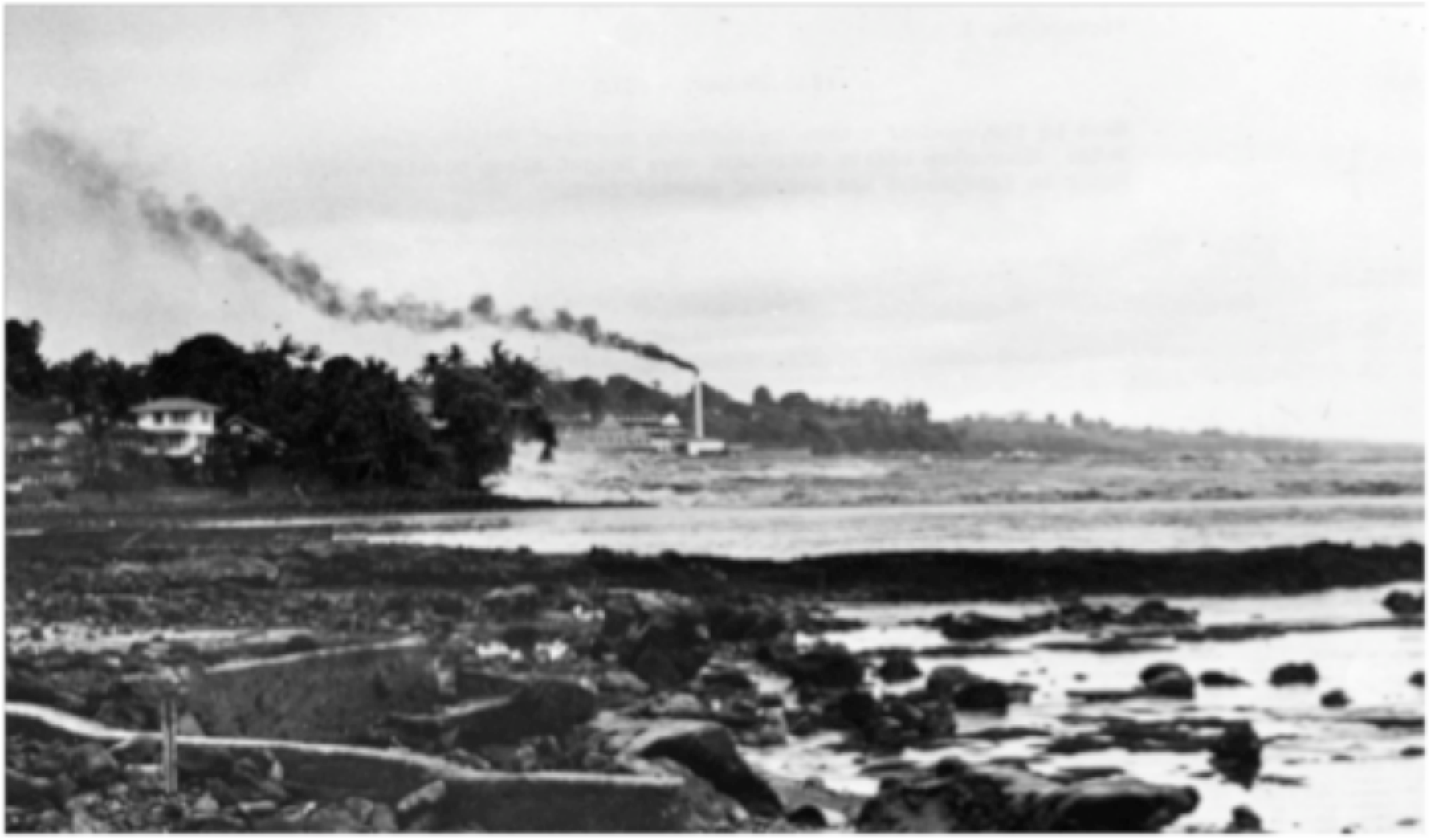}
\caption{Wave in the form of a bore, approaching Wailuku River, Hilo, during the 1946 tsunami. It appears to be moving along the seacliff as a Mach stem. (Courtesy Modern Camera Center, Hilo).}
\label{fig:Fig2}
\end{figure}

Wave refraction, reflection and diffraction lead to obliquely interacting waves in shallow water, and when their superposed amplitudes are sufficiently large, nonlinear effects can have striking effects on the resulting surface wave patterns. The nonlinear interaction of obliquely propagating solitary waves in similar directions can cause a high wave hump resulting from the crossing of the waves in the two-dimensional propagation space. Wiegel \cite{W:64} described that regular reflection of a solitary wave at a vertical rigid wall does not occur for sufficiently small angle of incidence, being replaced by a ÒMach reflectionÓ (a geometrically similar reflection from acoustics \cite{CF:48}). The apex of the incident and reflected wave separates away from the wall and is joined to it by a third solitary wave that perpendicularly intersects the wall as depicted in Figure \ref{fig:Fig1}. Wiegel \cite{W:64} further reported that when the 1946 Aleutian Tsunami hit the town of Hilo, Hawaii, a Mach-stem phenomenon was observed along the cliffs forming the western boundary of the Hilo Bay (Figure \ref{fig:Fig2}).

Here we focus on such shallow-water wave interactions: two identical solitary waves propagating with a small oblique angle 2$\psi_0$. This condition is equivalent to the Mach reflection with the incident wave angle $\psi_0$ to a perfectly reflective vertical wall. Miles \cite{M:77a},\cite{M:77b} presented a theoretical analysis of the Mach reflection at its asymptotic state. His theory predicts the extraordinary four-fold amplification of the stem wave. There were several laboratory and numerical experiments to attempt to validate Miles's four-fold amplification, but with no definitive success (for numerical experiments, e.g.\cite{F:80},\cite{T:93},\cite{KTK:98},\cite{BK:99}, and for laboratory experiments, \cite{P:57},\cite{M:80}).

In this paper, we reanalyze Mach reflection in a systematic fashion based on the Kadomtsev-Petviashvili (KP) equation. Our theoretical analysis allows us to predict not only the asymptotic states but also the developing states of the Mach reflection phenomenon. Then, we present the laboratory data obtained from the precision experiments, and quantitatively discuss agreement and discrepancies between the theoretical predictions and the laboratory results: also reanalyze the previous numerical results by Tanaka \cite{T:93}.

\section{The KP solitons}
\label{sec:1}
Based on the weakly nonlinear, weakly dispersive waves having the small oblique angles  $\psi_0^{2}=\mathcal O(\epsilon)$ with $0<\epsilon\ll 1$, we assume the following orders in the wave amplitude $a_0$ and the constant water depth $h_0$ at the quiescent state:
\[
\frac{a_0}{h_0}=\mathcal O(\epsilon),\qquad \left(\frac{h_0}{\lambda_0}\right)^2=\mathcal O(\epsilon),\qquad\tan^2 \psi_0=\mathcal O(\epsilon),
\]
where $\lambda_0$ is the wavelength scale. Then, the Euler formulation results in the Kadomtsev-Petviashvili (KP) equation of the form 
\begin{equation}\label{pKP}
\left(\eta_t+c_0\eta_x+\frac{3c_0}{2h_0}\eta\eta_x+\frac{c_0h_0^2}{6}\eta_{xxx}\right)_x+\frac{c_0}{2}\eta_{yy}=0,
\end{equation}
in which $\eta$ is the water-surface elevation from $h_0$, and $c_0= \sqrt{gh_0}$. An exact solution of the KP equation for one-line soliton can be written as:
\begin{equation}\label{pForm}
\eta = a_0{\rm sech}^2\sqrt{\frac{3a_0}{4h_0^3}}\left[ x+ y\tan\psi_0-c_0\left(1+\frac{a_0}{2h_0}+\frac{1}{2}\tan^2\psi_0\right) t\right].
\end{equation}
Except for the case $\psi_0=0$, the KP solution \eqref{pForm} does not satisfy the KdV equation; hence it does not support the KdV soliton.
One should also note that the solution \eqref{pForm} becomes unphysical for a large angle.
This is due to the assumption of the quasi-two dimensionality of the problem.  The effect of the finite angle can be
imposed in the following way:  first write the KP solution \eqref{pForm} in the form,
\begin{equation}\label{KP1}
\eta = a_0{\rm sech}^2\sqrt{\frac{3a_0}{4h_0^3\cos^2\psi_0}}\left[ x\cos\psi_0+ y\sin\psi_0-c_0\cos\psi_0 \left(1+\frac{a_0}{2h_0}+\frac{1}{2}\tan^2\psi_0\right) t\right].
\end{equation}
Define $\chi:=x\cos\psi_0+y\sin\psi_0$ (the normal direction to the one line-soliton wave crest), and note $\cos\psi_0=1/\sqrt{1+\tan^2\psi_0}=1-\frac{1}{2}\tan^2\psi_0+\mathcal{O}(\epsilon^2)$, we have
\begin{equation}\label{KP2}
\eta = a_0{\rm sech}^2\sqrt{\frac{3a_0}{4h_0^3\cos^2\psi_0}}\left[ \chi-c_0 \left(1+\frac{a_0}{2h_0}\right) t+\mathcal{O}(\epsilon^2)\right].
\end{equation}
Note here that the width of the soliton depends on the angle in the higher order, which is evidently unphysical. 
So we now define an equivalent $\eta$ including a higher order correction to the quasi-two dimensionality, 
\begin{equation}\label{Heta}
\hat\eta=\hat{a}_0{\rm sech}^2\sqrt{\frac{3\hat{a}_0}{4h_0^3}}\left[ \chi-c_0 \left(1+\frac{\hat{a}_0}{2h_0}\right) t+\mathcal{O}(\epsilon^2)\right], \quad {\rm with}\quad \hat{a}_0=\frac{a_0}{\cos^2\psi_0},
\end{equation}
which gives a better approximation than the KP solution \eqref{pForm} for the solitary wave with the propagation angle $\psi_0$ in physical situation.
Notice that with this new amplitude $\hat{a}_0$, the soliton solution \eqref{Heta} satisfies the KdV equation in the form
up to $\mathcal{O}(\epsilon^2)$,
\begin{equation}\label{KP3}
\hat\eta_t+c_0\hat\eta_{\chi}+\frac{3c_0}{2h_0}\hat\eta\hat\eta_{\chi}+\frac{c_0h_0^2}{6}\hat\eta_{\chi\chi\chi}=0.
\end{equation}
The point here is that the amplitude of the soliton observed in the experiment (or numerical simulations of the Euler
equation) should be represented by $\hat{a}_0$ instead of $a_0$ of \eqref{pForm}.

As for the KP limit, the solutions expressed by (\ref{pForm}) and (\ref{Heta}) are equivalent. None the less, it makes a significant difference when the theoretical predictions are compared with the data from laboratory and/or numerical experiments $-$ some of the experiments, including the results presented in this paper were performed with large amplitude $a_0/h_0$ and large incident angle $\psi_0$.

\subsection{Exact soliton solutions of the KP equation}\label{sec:1-1}
Let us write the KP equation  \eqref{pKP} in the following standard form,
\begin{equation}\label{sKP}
\left(4u_T+6uu_X+u_{XXX}\right)_X+3u_{YY}=0,
\end{equation}
where the non-dimensional variables $(X,Y,T)$ and $u$ are defined by
\begin{equation}\label{realvariables}
 \eta=\frac{2h_0}{3}\,u,\quad 
 x-c_0 t=h_0\,X,\quad  y=h_0\,Y,\quad  t=\frac{3h_0}{2c_0}\,T.
\end{equation}
Notice that the one-line soliton solution in the standard form  is given by
\begin{equation}\label{solution1}
u=A_0{\rm sech}^2\sqrt{\frac{A_0}{2}}\left(X+Y\tan\psi_0-CT\right),
\end{equation}
where $C=\frac{1}{2}A_0+\frac{3}{4}\tan^2\psi_0$ and the amplitude $A_0$ has the relation with the physical variables,
\begin{equation}\label{amp}
A_0=\frac{3a_0}{2h_0}=\frac{3\hat{a}_0}{2h_0}\,\cos^2\psi_0.
\end{equation}
With the standard form of the KP equation in \eqref{sKP}, we consider the solution in the form
\begin{equation}\label{u}
u(X,Y,T)=2\,[\ln\tau(X,Y,T)]_{XX}.
\end{equation}
where $\tau$ is referred to as the {\it tau} function defined in the Wronskian determinant,
\begin{equation}\label{tau}
\tau={\rm Wr}(f_1,f_2,\ldots,f_N).
\end{equation}
The functions $f_i$'s satisfy the linear equations, $f_Y=f_{XX}, f_T=-f_{XXX}$ (see e.g. \cite{H:04}).
For our soliton solutions, we choose the solution form of:
\begin{equation} 
f_i = \sum_{j=1}^{M} a_{ij}e^{\theta_j}, \quad
\theta_j = \kappa_j X + \kappa_j^2 Y - \kappa_j^3 T.
\label{eq:fis}
\end{equation}    
Here the coefficient matrix $A=(a_{ij})$ is a constant $N\times M$ matrix,
and $\kappa_j$ are constants with the ordering $\kappa_1 < 
\cdots < \kappa_M$. Thus each solution $u(X,Y,T)$ is then completely determined by
the $A$-matrix and the $\kappa$-parameters.

Single soliton solution is obtained in the case with $N=1$ and $M=2$:
We have $\tau = e^{\theta_i} + e^{\theta_j} $  for $\kappa_i<\kappa_j$, and $u=2(\ln\tau)_{XX}$ gives
\begin{equation} 
u = \frac{1}{2}(\kappa_j-\kappa_i)^2\mbox{sech}^2\frac{1}{2}(\theta_j-\theta_i).
\label{eq:1-soliton-sol}
\end{equation}
The line $\theta_i=\theta_j$ represents the ridge of the soliton, 
and we refer to this {\it line}-soliton as $[i,j]$-soliton with $i<j$ (i.e. $\kappa_i<\kappa_j$). 
The amplitude $A_{[i,j]}$ and the inclination $\tan\psi_{[i,j]}$ of this 
soliton are given by 
\begin{equation} 
A_{[i,j]} = \frac{1}{2}(\kappa_j-\kappa_i)^2, \quad  
\tan\psi_{[i,j]} = \kappa_i + \kappa_j,   
\label{eq:1-soliton-amp-dir}
\end{equation} 
where $-\pi/2 < \psi_{[i,j] }< \pi/2$ is the angle of the line soliton
 measured counterclockwise from the $Y$-axis. This angle also represents 
the propagation direction of the line soliton (see Figure\ref{fig:Fig3}).

\subsection{The Mach reflection and the KP solutions}\label{sec:1-2}
For the Mach reflection problem, we take $N=2, M=4$. Using the Binet-Cauchy theorem for the determinant,
the $\tau$-function  $\tau={\rm Wr}(f_1,f_2)$ 
can be expressed in the form,
\begin{equation}
\tau = \sum_{1 \le i<j \le 4}\xi(i,j) E(i,j),
\label{eq:tau-function3}
\end{equation}
where $\xi(i,j)$ is the $2 \times 2$ minor consisting of   
$i$-th and $j$-th columns of the $A$-matrix, and $E(i,j)={\rm Wr}(e^{\theta_i},e^{\theta_j})=(\kappa_j-\kappa_i)e^{\theta_i+\theta_j}$. 
For the regular solutions, we require all of these minors 
to be non-negative (note $E(i,j)>0$ with the order $\kappa_i<\kappa_j$).

\begin{figure}[h]
\begin{center}
 \includegraphics[width=42.0mm]{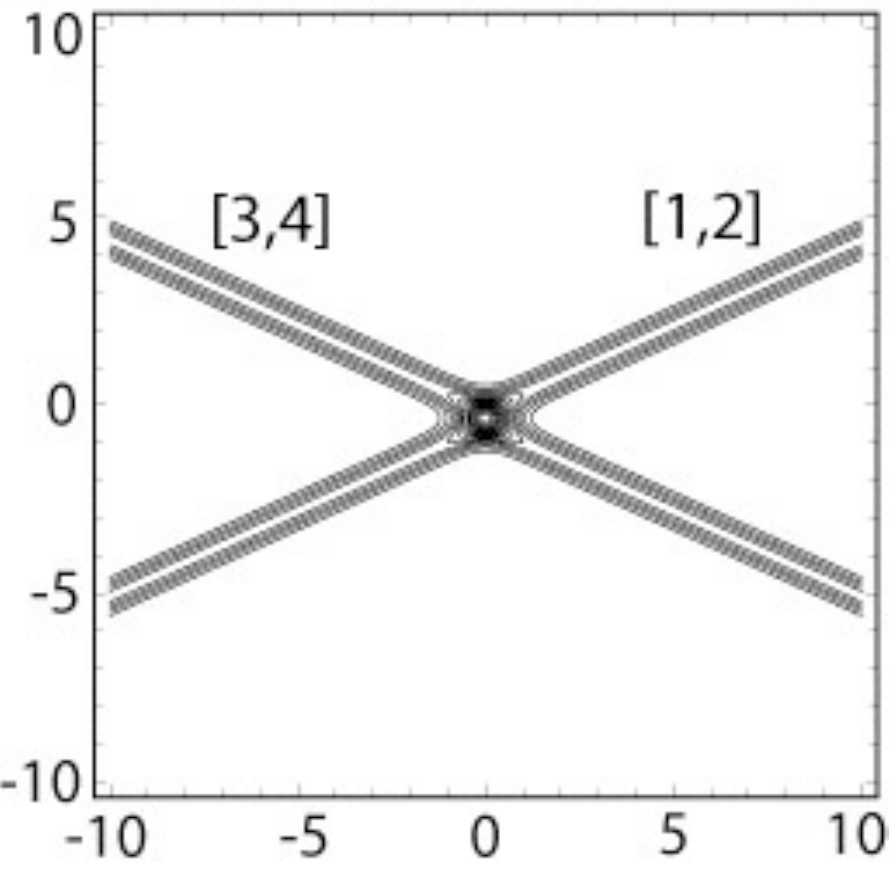}\hskip 2.1cm
 \includegraphics[width=42.0mm]{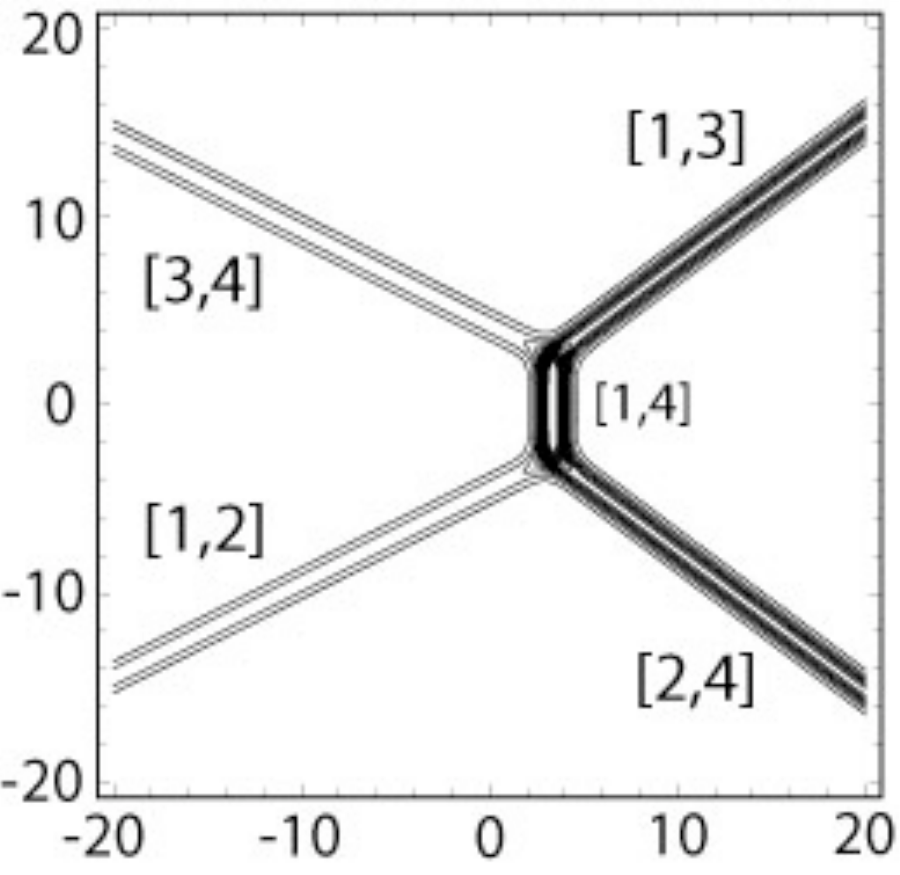}\\
 \bigskip
 \hskip 0.2cm \includegraphics[width=42.0mm]{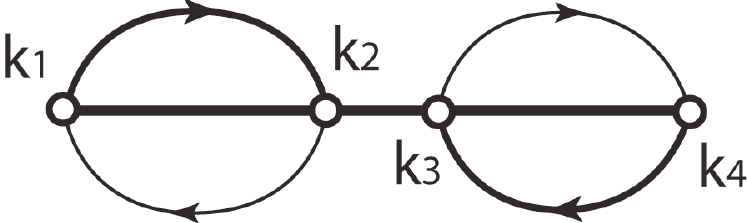}\hskip 2.4cm
\raisebox{-2.0mm}{ \includegraphics[width=35.0mm]{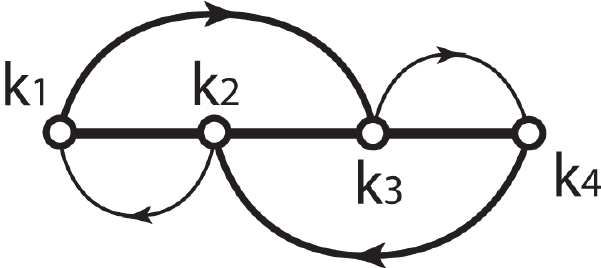}}
 \caption{Contour plots and the chord diagrams of soliton solutions. Left: O-type 
 solution. Right: (3142)-type solution. Each $[i,j]$ 
denotes the $[i,j]$-soliton. The length of $[1,4]$-soliton changes in $t$.
The upper (lower) chords represent the asymptotic solitons in $Y\gg0~(Y\ll0)$.
The thicker chords correspond to the solitons in the right side ($X\gg0$),
and the arrows on the chords show the pairings in the permutations.}
 \label{fig:Fig3}
 \end{center}
\end{figure}

As was shown in \cite{CK:08},\cite{CK:09}, each $\tau$-function (\ref{eq:tau-function3}) 
generates a soliton solution which consists of at most
two line-solitons for both $y \to \pm\infty$. We consider the following two types which are relevant to the solutions
 of the initial value problems for the Mach reflection:
 one consists of  two line-solitons of $[1,2]$ and $[3,4]$
for both $|Y|\gg 0$, which is called O-type soliton (``O'' stands for {\it original}, see \cite{K:04},\cite{CK:09}); the other one consists of
$[1,3]$ and $[3,4]$ line-solitons for $Y \gg 0$ and 
$[1,2]$ and $[2,4]$ line-solitons for $Y \ll 0$. 
Let us call this soliton $(3142)$-type, because those four line-solitons
represent a permutation $\pi=\binom{1~2~3~4}{3~1~4~2}$.
It was shown in general (see \cite{CK:08},\cite{CK:09}) that each exact solution generated by the Wronskian determinant with \eqref{eq:fis}
can be parametrized by a unique element of the permutation group.
In this representation, O-type is expressed as $(2143)$-type solution.
Figure \ref{fig:Fig3} illustrates the contour plots of O-type and (3142)-type solutions in the $XY$-plane, and the corresponding
{\it chord diagrams} which represent each soliton as a chord
joining a pair of $\kappa_i$'s following its permutation representation. The upper chords represent the asymptotic solitons $[i,j]$ for $Y\gg 0$, and the lower chords for the asymptotic solitons $[i,j]$ for $Y
\ll 0$.  Note also that the length and the location of each chord give the amplitude and
the angle of the corresponding soliton (cf. (\ref{eq:1-soliton-amp-dir})).
The $A$-matrices for those solutions are respectively given by 
\begin{equation}
A_{\rm O}=
\begin{pmatrix}
1 & a & 0 & 0 \cr
0 & 0 & 1 & b\cr
\end{pmatrix}
, \qquad A_{(3142)}=
\begin{pmatrix}
1 & a & 0 & -c \cr
0 & 0 & 1 & b\cr
\end{pmatrix},  
\label{eq:O-3142matrix}
\end{equation} 
where $a, b, c > 0$ are constants determining the locations of the solitons (see \cite{CK:09}). Notice that the $\tau$-function  for the (3142)-type contains {\it five} exponential terms,
\begin{equation}
\begin{array}{llll}
\tau \!=&\displaystyle{\!(\kappa_3\!-\!\kappa_1)\,e^{\theta_1+\theta_3} 
 \!+\! (\kappa_4\!-\!\kappa_1)\,b\,e^{\theta_1+\theta_4} 
 \!+\! (\kappa_3\!-\!\kappa_2)\,a\,e^{\theta_2+\theta_3}} \\[1.5ex]
 \!&\displaystyle{+\! (\kappa_4\!-\!\kappa_2)\,a\,b\,e^{\theta_2+\theta_4} \!+\! (\kappa_4\!-\!\kappa_3)\,
c\,e^{\theta_3+\theta_4}},\end{array}
\label{eq:3142-tau}
\end{equation}
and the $\tau$-function for O-type with $c=0$ in (\ref{eq:3142-tau}) contains only {\it four} terms.

Let us fix the amplitudes and the angles of the solitons in the positive $X$ regions
for both O- and (3142)-types, so that those solutions are symmetric with respect to
the $X$-axis (see Figure \ref{fig:Fig3}):
\begin{equation}
\begin{array}{llll}
A_0&\equiv\left\{\begin{array}{l}A_{[1,2]}=A_{[3,4]} \quad {\rm (for~O-type)}\\[1.0ex]
A_{[1,3]}=A_{[2,4]}\quad {\rm( for~(3142)-type)}\end{array}\right.\\
\\
\psi_0&\equiv\left\{\begin{array}{l}-\psi_{[1,2]}=\psi_{[3,4]}>0\quad {\rm (for~O-type)}\\[1.0ex]
-\psi_{[1,3]}=\psi_{[2,4]}>0\quad {\rm (for~(3142)-type)}\end{array}\right.
\end{array}
\end{equation}
Then from (\ref{eq:1-soliton-amp-dir}),
one can find the $\kappa$-parameters  in terms of $A_0$ and $\tan\psi_0$
with $\kappa_1=-\kappa_4$ and $\kappa_2=-\kappa_3$ (due to the symmetry):
In the case of O-type, we have 
\begin{equation}
\kappa_1=-\frac{1}{2}\left(\tan\psi_0+\sqrt{2A_0}\right), \quad 
\kappa_2=-\frac{1}{2}\left(\tan\psi_0-\sqrt{2A_0}\right).
\label{eq:values of k (O)}
\end{equation}  
The ordering 
$\kappa_2<\kappa_3$ then implies $\tan\psi_0 > \sqrt{2A_0}$. 
On the other hand, for the (3142)-type, we have    
\begin{equation}
\kappa_1=-\frac{1}{2}\left(\tan\psi_0+\sqrt{2A_0}\right), \quad 
\kappa_2=\frac{1}{2}\left(\tan\psi_0-\sqrt{2A_0}\right).
\label{eq:values of k}
\end{equation} 
The ordering $\kappa_2<\kappa_3$  implies
$\tan\psi_0<\sqrt{2A_0}$.

Thus, if all the solitons in the positive $X$-region have the same amplitude $A_0$
for both O- and (3142)-types, then an O-type solution arises when $\tan\psi_0>\sqrt{2A_0}$,
and a (3142)-type when $\tan\psi_0<\sqrt{2A_0}$. 
Then the limiting value at $\kappa_2=\kappa_3$ $(=0)$
 defines the critical angle $\psi_c$,
\begin{equation}\label{critical}
\tan\psi_c:=\sqrt{2A_0}.
\end{equation}
Note from (\ref{eq:3142-tau}) that at the critical angle, i.e. $\kappa_2=\kappa_3$, the $\tau$-function
has only {\it three} exponential terms, and this gives a ``Y"-shape resonant solution
as Miles noted \cite{M:77b}.

One can also see from (\ref{eq:1-soliton-amp-dir}) that for (3142)-type solution, the solitons
in the negative $X$-region are smaller than those in the positive region, i.e.
$A_{[3,4]}=A_{[1,2]}=\frac{1}{2}\tan^2\psi_0 < A_0$, and the angles of those in the negative $x$-regions
do not depend on $\psi_0$ and
$\psi_{[3,4]}=-\psi_{[1,2]}=\psi_c$. 
Two sets of three solitons $\{[1,3],[1,4],[3,4]\}$ 
and $\{[2,4],[1,4],[1,2]\}$ are both in the soliton 
resonant state. These properties of the (3142)-type solution 
are the same as those of Miles's asymptotic solution 
for the Mach reflection of a shallow water 
soliton \cite{M:77b}.   However one should note that the $[1,4]$-soliton corresponding to the
Mach stem becomes a soliton solution only in an asymptotic sense.
Then the exact solution of $(3142)$-type can provide an estimate of a propagation distance, at which the amplitude is sufficiently developed.
\begin{figure}[h]
\centering
\includegraphics[scale=0.45]{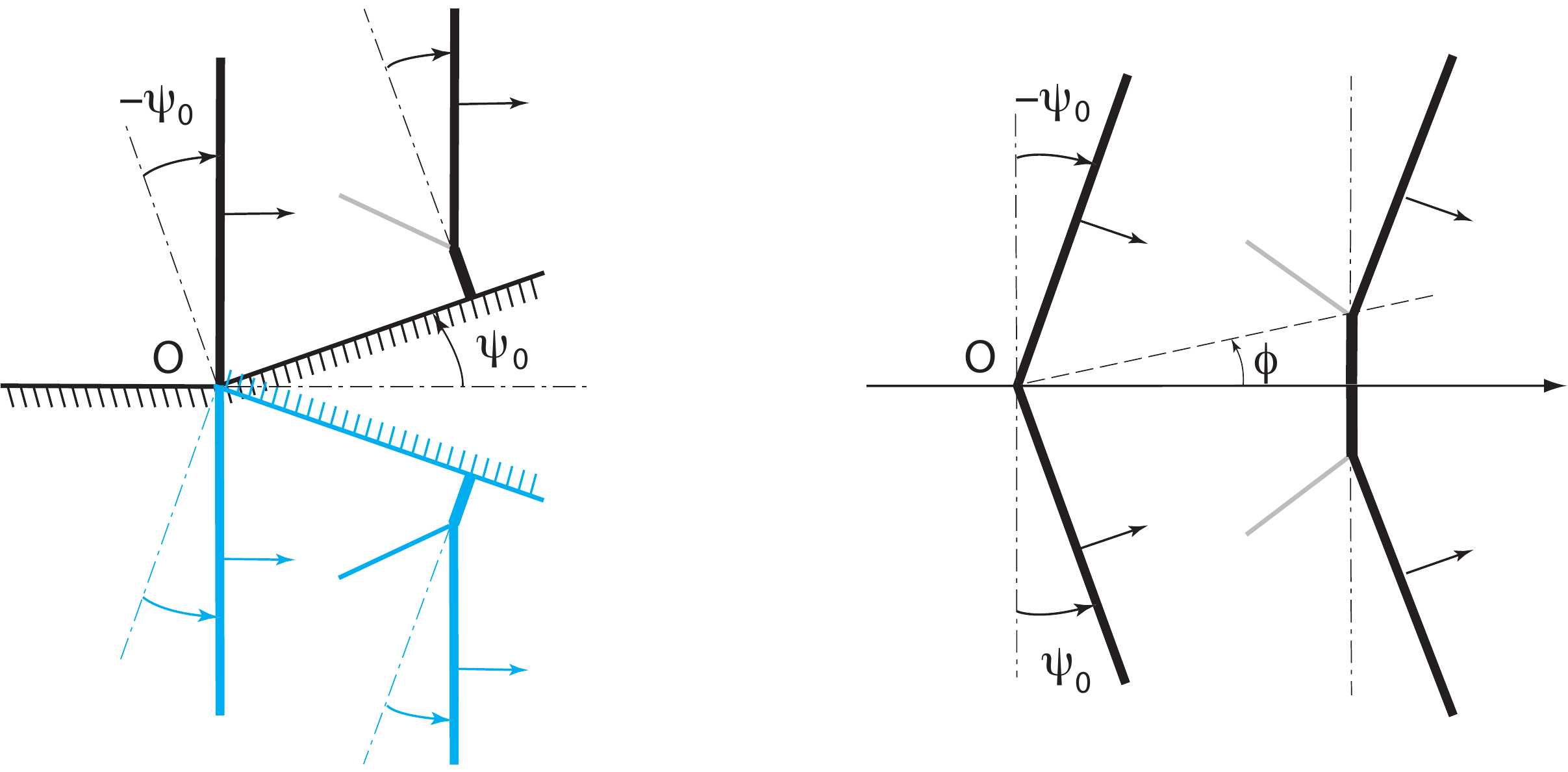} 
\caption{The Mach reflection. The left figure illustrates a semi-infinite line-soliton propagating
parallel to the wall. The right figure is an equivalent system to the left one when we ignore
the viscous effect on the wall. The resulting wave pattern is of $(3142)$-soliton solution.
The angle $\phi$ becomes zero if the initial angle satisfies $\psi_0\ge \psi_c$, i.e. no stem.}
\label{fig:Fig4}
\end{figure}

\section{The Mach reflection in shallow water waves}\label{sec:2}
Miles \cite{M:77a},\cite{M:77b} considered an oblique interaction of two line-solitons  using O-type solutions.
He observed that resonance occurs at the critical angle $\psi_c$, and when the angle $\psi_0$
is smaller than $\psi_c$, the O-type solution becomes
singular (recall that at the critical angle $\psi_c$, one of the exponential term in the $\tau$-function vanishes). 
He also noticed a similarity between this resonant interaction and the Mach reflection. This may be illustrated by the left figure of Figure \ref{fig:Fig4},
where an incidence wave shown by the vertical line is propagating to the right, and it hits a rigid wall
with the angle $-\psi_0$: the incidence wave angle $\psi_0$ is measured counterclockwise from the axis perpendicular to the wall
(see also \cite{F:80}).
If the angle of the incidence wave (equivalently the inclination angle of  the wall) is large, 
the reflected wave behind the incidence
wave has the same angle $\psi_0$, i.e. a regular reflection occurs. However, if the angle is small,
then an intermediate wave called the Mach stem appears as shown in Figure \ref{fig:Fig4}.
The critical angle for the Mach reflection is given by the angle $\psi_c$.
The Mach stem, the incident wave and the reflected wave interact resonantly, and
those three waves form
a resonant triplet. The right one in Figure \ref{fig:Fig4} illustrates an equivalent system of
the wave propagation in the left figure, (ignoring the effect of viscosity on the wall, i.e. no boundary layer). At the point $O$, the initial wave has V-shape
with the angle $\psi_0$, which forms the initial data for the simulation.
Then the numerical simulation describes the reflection 
of line-soliton with an inclined wall, and these results explain well the appearance of the Mach
reflection in terms of the exact soliton solution of $(3142)$-type (see \cite{KOT:09},\cite{KK:10}).

The maximum amplitude for this problem occurs at the wall
and this can be obtained by the following formula (see \cite{M:77b},\cite{CK:09}):   For the O-type solution with 
$\tan\psi_0>\tan\psi_c=\sqrt{2A_0}$, 
\begin{equation}\label{Omax}
u_{\rm max}=\frac{4A_0}{1+\sqrt{1-k^{-2}}}, \qquad
{\rm with}\quad k:=\frac{\tan\psi_0}{\sqrt{2A_0}}>1,
\end{equation}
and  for $(3142)$-type solution when $k<1$,
\begin{equation}\label{stemA}
A_{[1,4]}=\frac{1}{2}(\kappa_4-\kappa_1)^2=A_0\left(1+k\right)^2.
\end{equation}
Note here that at the critical angle $\psi_0=\psi_c$ (i.e. $k=1$), the amplitude takes the maximum $u_{\rm max}=A_{[1,4]}=4A_0$ (i.e. $\alpha_w=u_{\rm max}/A_0=A_{[1,4]}/A_0=4$).


\section{Laboratory Experiments}\label{sec:3}
\begin{figure}[h]
\centering
\includegraphics[width=4.5in,height=2.45in]{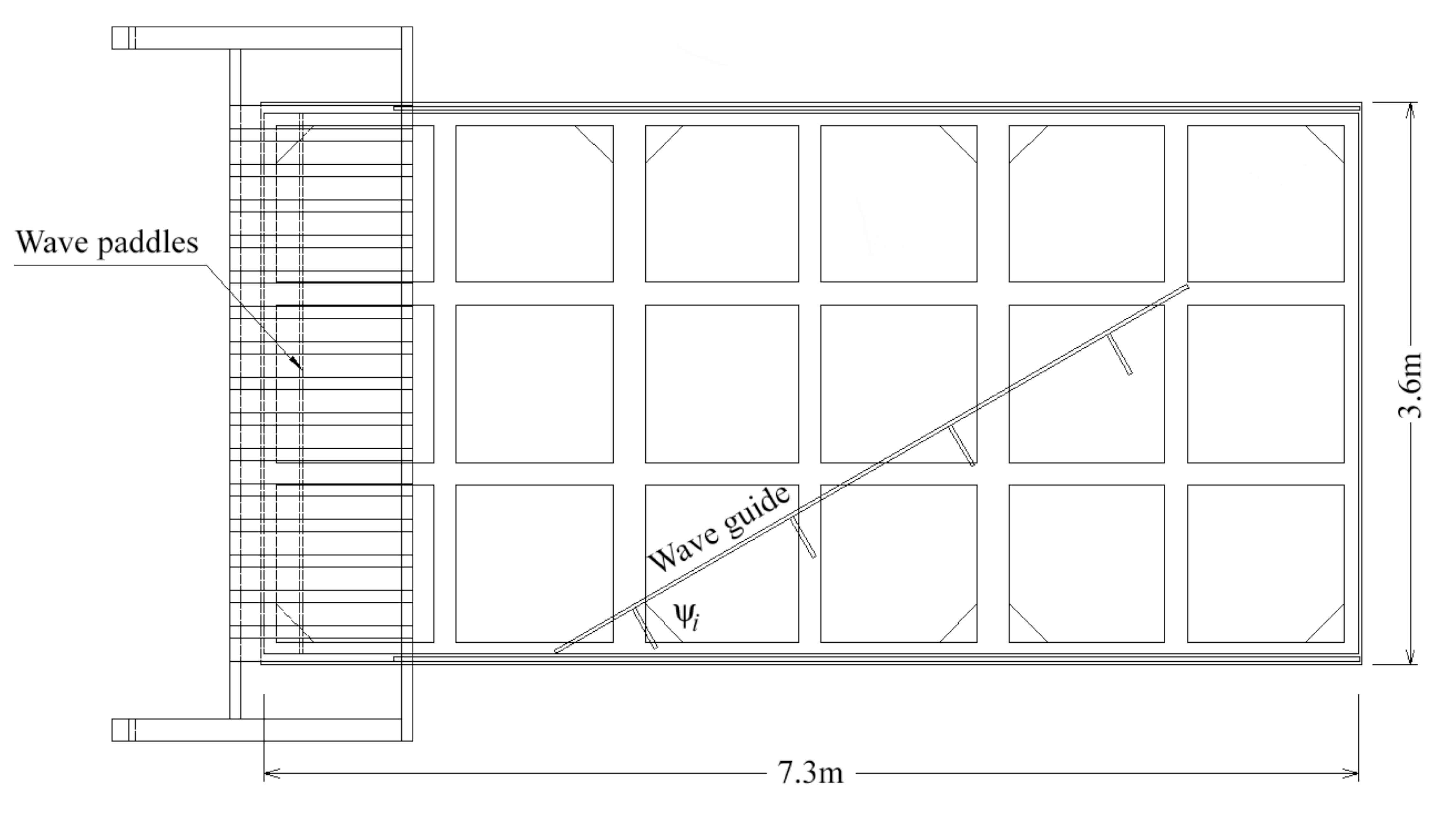}
\includegraphics[width=4.5in,height=1.45in]{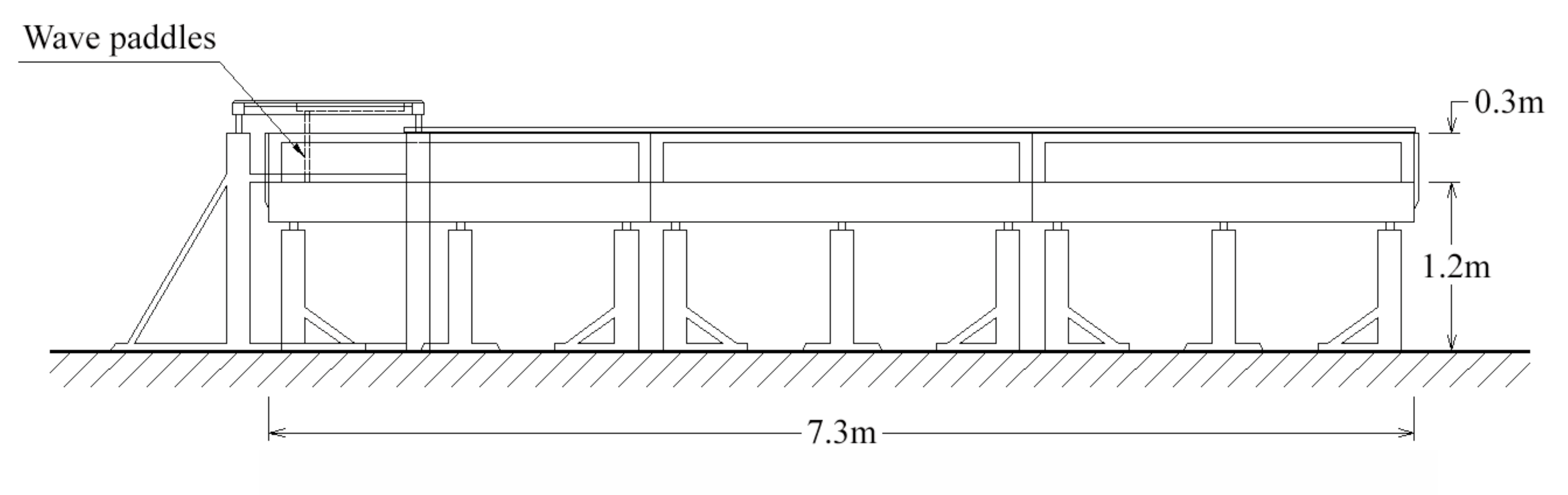}
\caption{Schematic drawings of the laboratory apparatus: (a) a plan view showing the waveguide that creates an oblique wave reflection and the tank frame that can be seen through the bed that is made of glass plates; (b) an elevation view.}\label{fig:Fig5}
\end{figure}

Laboratory experiments were performed in a wave tank (7.3 m long, 3.6 m wide, and 0.30 m deep). The wave basin is equipped with a 16-axis directional-wave maker system along the 3.6-m long sidewall, capable of generating arbitrary-shaped, multi-directional waves. The paddles are made of PVDF (Polyvinylidene fluoride) plates that are moved horizontally in piston-like motions. Each paddle has a maximum horizontal stoke of 55cm Ð more than adequate to generate very long waves with a water depth of 6.0 cm. While the wavemaker is capable of generating arbitrary-shaped, multi-directional waves, we chose to generate a solitary wave in the normal direction (along the sidewalls). An obliquely incident solitary wave was created by placing a 2.54 cm thick Plexiglas plate vertically at a prescribed azimuth angle from the tankÕs sidewall (see Figure \ref{fig:Fig5}). Because the wave paddles are driven synchronously along the sidewalls, any ambiguity associated with potential deviation caused by the paddle deformation is eliminated. Solitary waves are generated using the algorithm based on KdV solitons \cite{G:79}. 
\begin{figure}[t]
\centering
\includegraphics[scale=0.6]{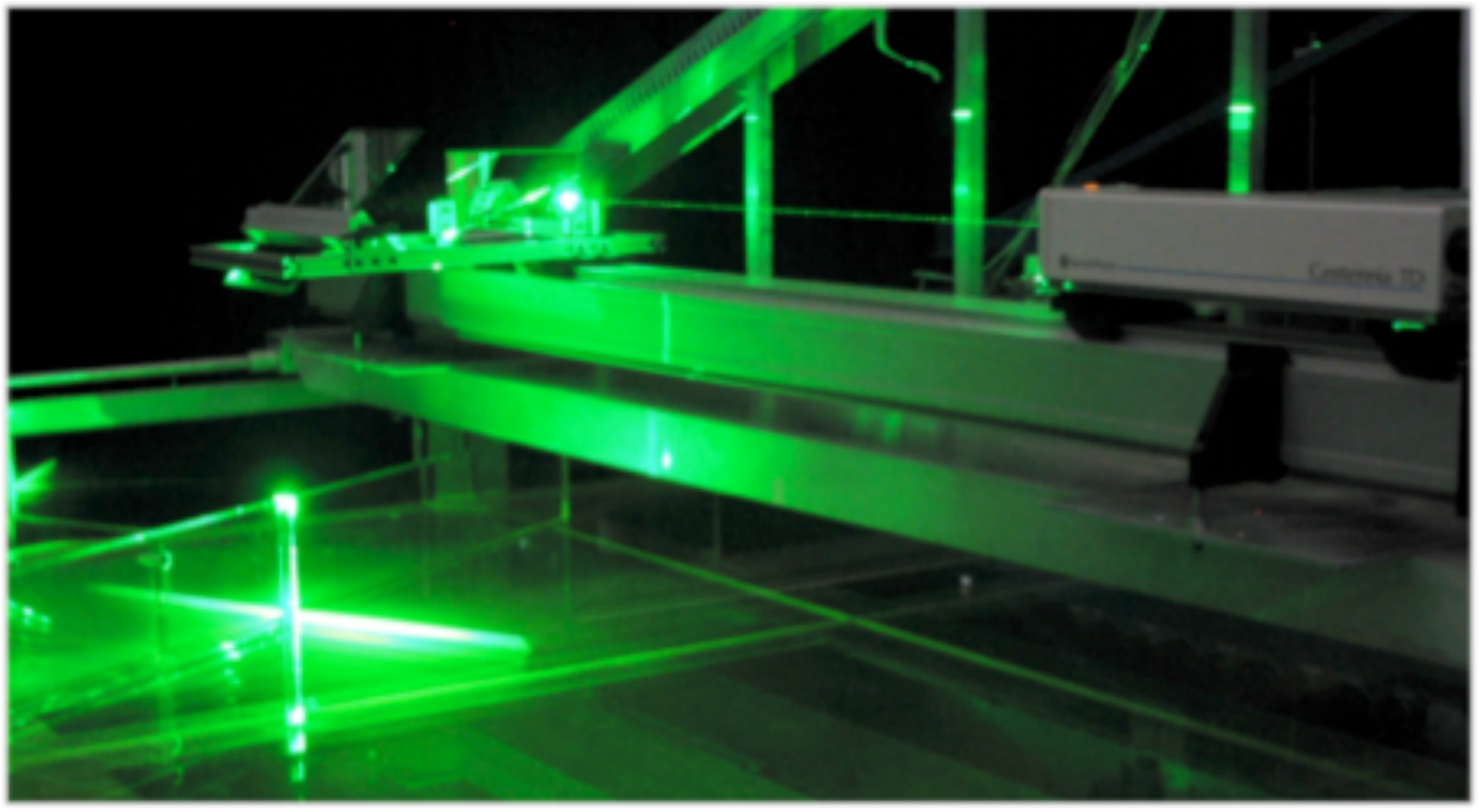} 
\caption{Setup for the Laser-Induced Fluorescent (LIF) method, comprised of the 5W laser, the cylindrical lens, and two front-face mirrors. The water dyed with fluorescein (green) fluoresces when excited by the laser sheet.}\label{fig:Fig6}
\end{figure}
\begin{figure}[h]
\centering
\subfloat[]{\includegraphics[width=2.2in,height=2.2in]{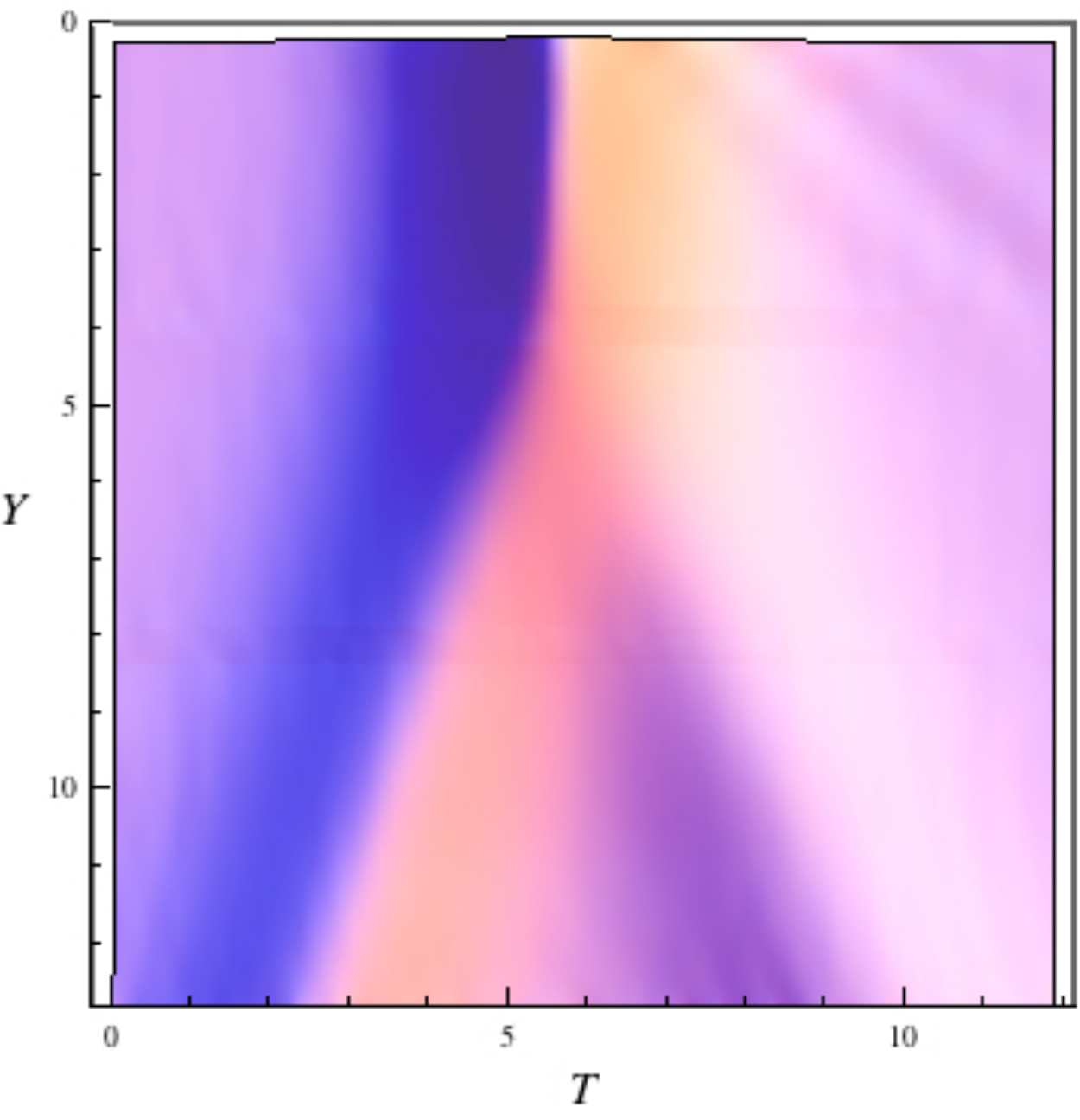} \hskip 1.0cm \includegraphics[width=2.2in,height=1.9in]{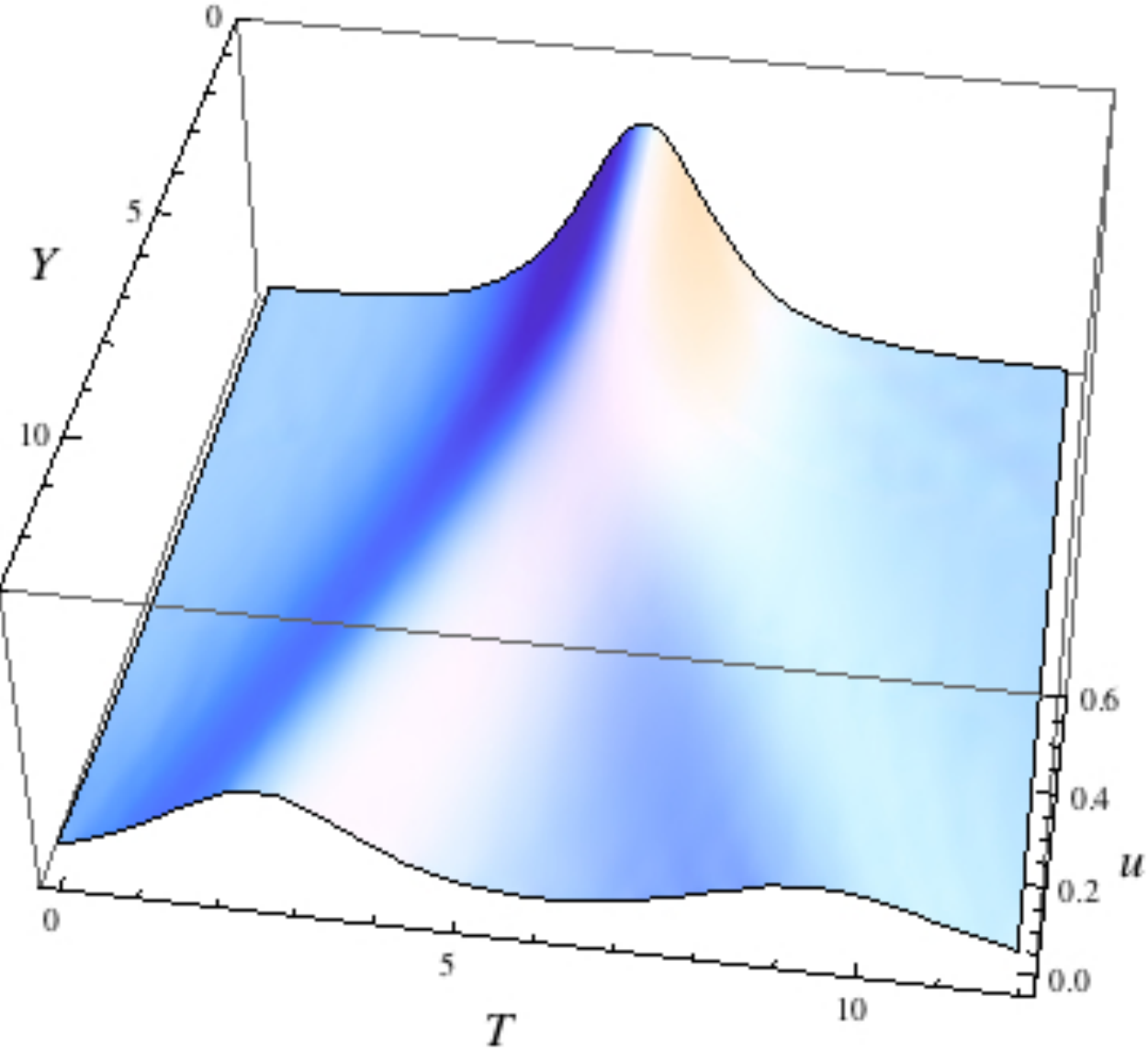}
\label{fig:Fig7a}}\\
\subfloat[]{\includegraphics[width=2.2in,height=2.2in]{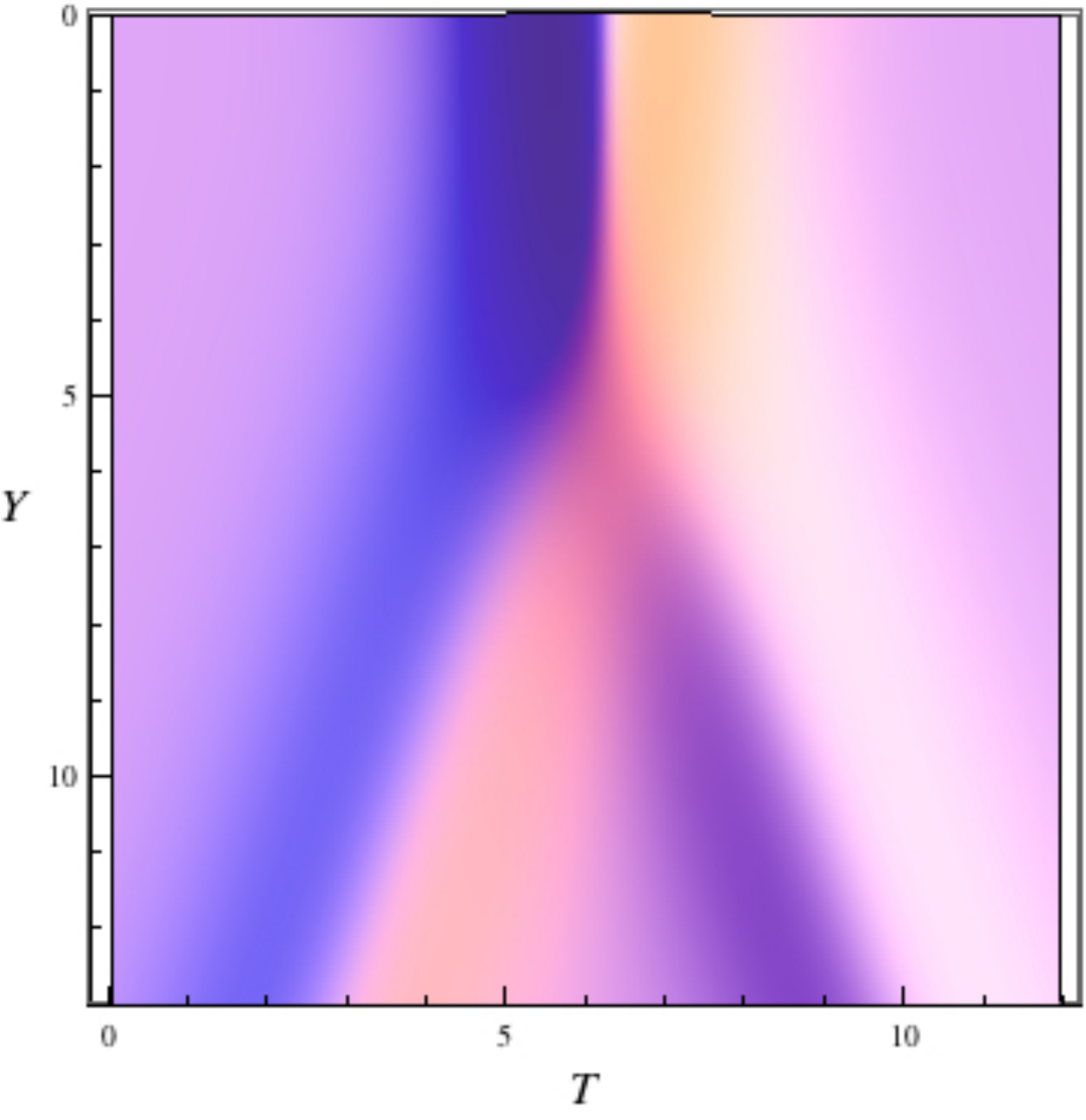} \hskip 1.0cm \includegraphics[width=2.2in,height=1.9in]{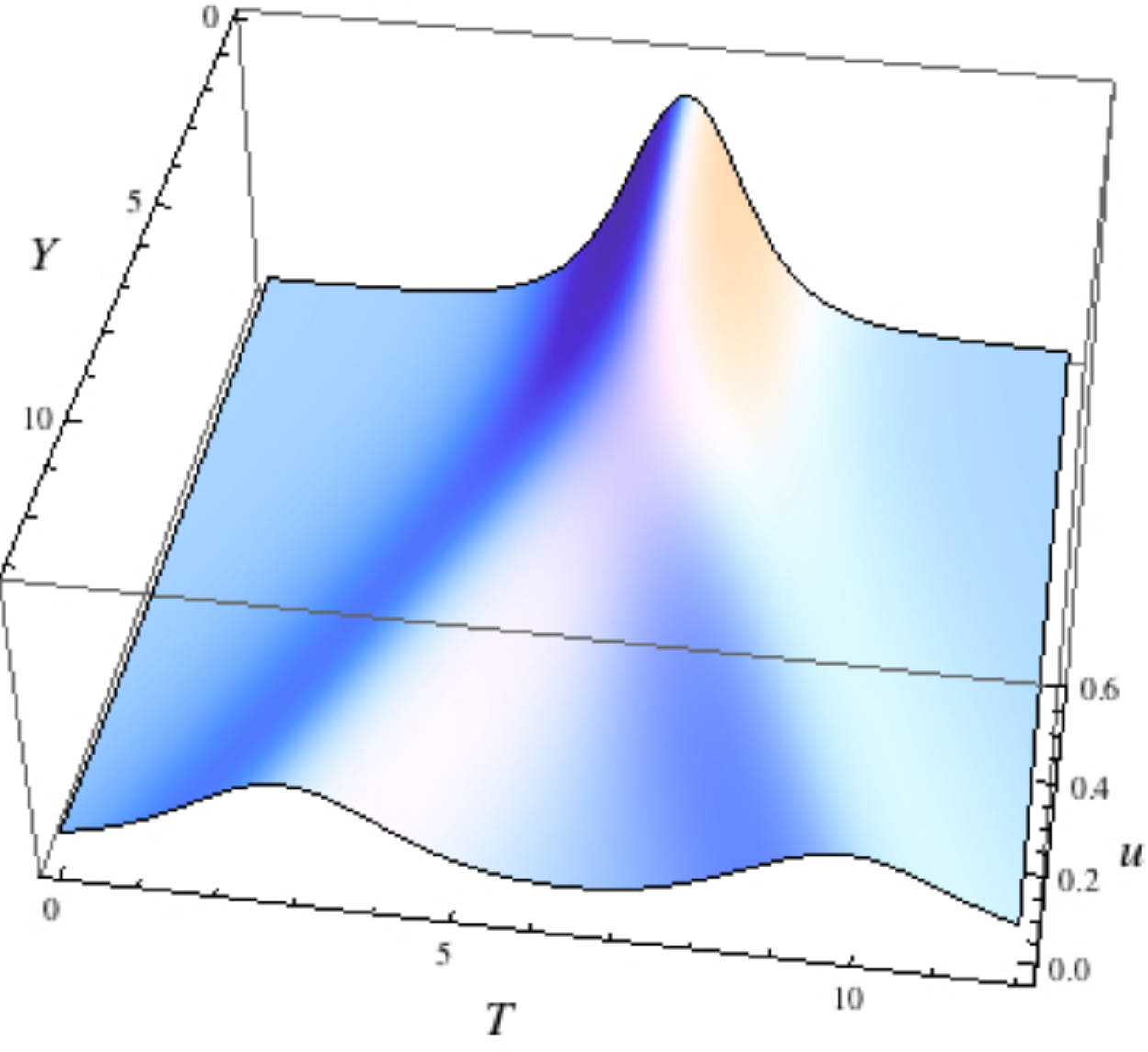} 
\label{fig:Fig7b}}
\caption{Two views of the temporal variation of the water-surface profile in the $Y$-direction (perpendicular to the wall) at $X = 71.1$: (a) the experimental result, and (b) the
corresponding $(3142)$-type exact soliton solution of the KP equation. The incident wave amplitude $A_0 = 0.212$, and the angle $\psi_0 = 30^{\circ}$. The data were obtained by the LIF method with the vertical laser sheet along the $Y$-direction.}\label{fig:Fig7}
\end{figure}

To examine temporal and spatial variations of water-surface profiles, we used the Laser Induced Fluorescent (LIF) method. Figure \ref{fig:Fig6} shows a setup for the LIF method used in this study. A laser beam (a 5W diode-pumped solid-state laser head mounted on the traversable carriage) is converted to a thin laser sheet using a cylindrical lens. Two front surface mirrors direct the laser sheet to illuminate the vertical plane perpendicular to the wall. With the aid of fluorescein dye dissolved in water, the vertical laser-sheet illumination from above induces the dyed water to fluoresce and identifies the water-surface profile directly and non-intrusively. The illuminated images are recorded by a high-speed high-resolution video camera (1280 $\times$ 1024 pixels and 100 frames per sec.). The captured images are rectified and processed with the calibrated image so that the resulting images can be analyzed quantitatively. Consequently, the LIF method is a highly accurate measurement technique to capture the spatiotemporal water-surface profiles in the non-intrusive manner.

The origin of coordinates is taken at the edge of the oblique vertical wall; x-direction points horizontally along the wall; y-direction points perpendicularly away from the wall; and z-direction points upward. The present experiments focus on the conditions of the quiescent water depth $h_0$ = 6.0 cm and the incident wave angles $\psi_0 = 20^{\circ}, 30^{\circ}$ and $40^{\circ}$. Hereinafter, we present the data in the nondimensionalized forms consistent with (\ref{realvariables}) unless stated otherwise.

Compiling three LIF segments of the water-surface profiles in the $Y$-direction yields a water-surface map in the $YT$-plane with adequate coverage, $0 <Y< 13.0$, as shown in Figure \ref{fig:Fig7}a. The figure represents the wave profile at $X = 71.1$ for the case of the incident wave amplitude $A_0 = 0.212$ with the oblique wall $\psi_0=30^{\circ}$. The plot in Figure \ref{fig:Fig7}a is made from 250 slices of the spatial profiles $-$ 100 slices per second $-$ with approximately 3000-pixel resolution in the $Y$-direction. As expected, the stem-wave formation is realized, in which the apices of the incident and reflected waves separate away from the wall by the third wave that perpendicularly intersects the wall. Figure \ref{fig:Fig7}b shows  the corresponding $(3142)$-type exact solution of the KP equation. For $A_0 = 0.212$ and $\psi_0=30^{\circ}$, we found from (\ref{eq:1-soliton-amp-dir}) that $\kappa_i=(-0.614, -0.037, 0.037, 0.614)$, and the $A$-matrix (\ref{eq:O-3142matrix}) is given by
\[
A=
\begin{pmatrix}
1 & 8.797 & 0 & -1.128\cr
0 & 0 & 1 & 0.530\cr
\end{pmatrix}.
\]
This choice of the $A$-matrix places the incidence wave crossing at the origin $(X=0,Y=0)$ at $T=0$ \cite{CK:09}.
The resulting wave profile shown in Figure \ref{fig:Fig7}b  is in excellent agreement with the laboratory observation, although a careful observation reveals that the measured stem-wave amplitude is slightly lower and the reflected wave amplitude is also lower than the prediction. This is because the wave reflection in the laboratory is still in the process of being established, while the reflected wave is assigned a priori as the $[3,4]$-soliton in the KP theory.

\begin{figure}[t]
\centering
\includegraphics[width=4in,height=3.0in]{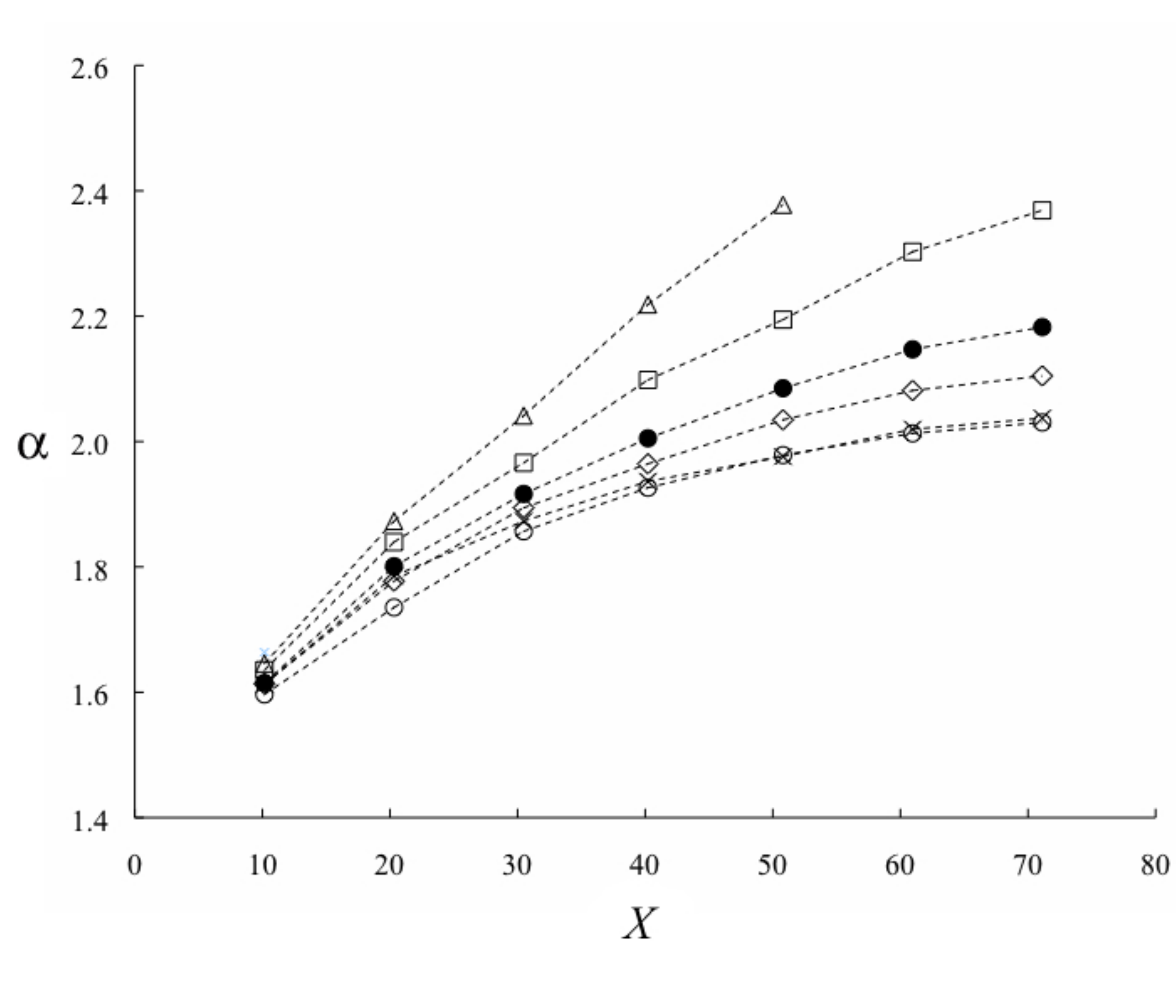}
\caption{Amplification growth $\alpha$ of the stem wave for the conditions of $\psi_0$ = $30^{\circ}$, and  - -$\times$ - -, $A_0$ = 0.086; - - $\circ$ - -, $A_0$ = 0.108; - - $\Diamond$ - -, $A_0$ = 0.161; - - $\bullet$ - -, $A_0$ = 0.212; - - $\square$ - -, $A_0$ = 0.312; - - $\triangle$ - -, $A_0$ = 0.413.} \label{fig:Fig8}
\end{figure}

Growing stem-wave amplification factors $\alpha$ $(= A_{[1,4]}/A_0)$ induced by a variety of incident waves with amplitudes $0.086 < A_0 < 0.413$ and $\psi_0=30^{\circ}$ are presented in Figure \ref{fig:Fig8}. No complete data are presented for the case of $A_0 = 0.413$ because of the wave breaking after $X=50.8$. 

The stem amplification continues to grow, and it is evident that the limited physical dimension of the laboratory apparatus prevents the Mach stem from reaching its fully developed state.  In Table \ref{Table1}, the experimental results of the amplification factor $\alpha$ are compared with those KP exact solutions (i.e. O-type for $k>1$ and $(3142)$-type for
$k<1$).  At $X=71.1$, the farthest measuring location in our experiments, the experimental results are in good agreement with those calculated from the KP exact solutions.
\begin{table}[h]
\centering
\caption{Amplification factor $\alpha$ of the stem waves for different $A_0$  
($k=\tan\psi_0/\sqrt{2A_0}$) with $\psi_0=30^{\circ}$:
$\alpha_{X=71.1}$(Exp.) are the laboratory data at $X=71.1$ and
$\alpha_{X=71.1}({\rm KP})$ are calculated from the corresponding KP exact solutions.
$\alpha_{X=\infty}$(Exp.) are estimated from the exponential curve fit to the laboratory data, and $\alpha_{X=\infty}$(KP) are
from \eqref{Omax} and \eqref{stemA}.  In the last row of $A_0=0.413$, the values of $\alpha$ in the brackets are obtained at
 $X = 50.8$, because of the wave breaking immediately after this point; hence, the greater amplification cannot be realized.   
 Notice the large deviations in the estimates $\alpha_{X=\infty}$(Exp.) in the boxes from the theoretical predictions for the cases near $k=1$.\label{Table1}}
\begin{tabular}{ccccccc}
\hline
\multicolumn{1}{c}{$A_0$} &\multicolumn{1}{c}{$k$}& \multicolumn{1}{c}{$\alpha_{X=71.1} ({\rm Exp.})$} 
& \multicolumn{1}{c}{$\alpha_{X=71.1}({\rm KP})$} &
 \multicolumn{1}{c}{$\alpha_{X=\infty} ({\rm Exp.})$} & \multicolumn{1}{c}{$\alpha_{X = \infty}({\rm KP})$}\\
\hline
0.086 & 1.392& 2.10 & 2.36&  2.13 & 2.36 \\
0.108 & 1.242 & 2.13 &  2.51 &          2.19 & 2.51\\
0.161 & 1.017& 2.24 &    3.38 &     \fbox{2.33} & \fbox{3.38} \\
0.212 & 0.887& 2.33 &    2.43 &     \fbox{2.46} & \fbox{3.56} \\
0.312 & 0.731& 2.52 &    2.61 &        2.92 &2.99 \\
0.413 & 0.635&   (2.48)     &     (2.54)          &    3.94 &   2.67 & \\
\hline
\end{tabular}
\end{table}
We estimate the asymptotic amplifications using the exponential curve fitting to the data: $\alpha = a e^{-bx}+ c$ , where $a$, $b$ and $c$ are positive constants to be determined. Those estimates also agree well, except for the case near $k=1$ (i.e. $A_0=0.161$ and $0.212$). Near $k=1$, the discrepancy is significant and the theory predicts approximately $45\%$ greater than the amplification projected from the experimental results. This suggests that while the growth rate be reduced gradually, the theoretical growth must be sustained for a much longer time than the actual growth in the real (viscous) fluid environment.

The present laboratory data are also compared with the data from Tanaka's  numerical experiments \cite{T:93}: his experiments are based on the high-order spectral method. This higher-order model allows him to study conditions less restricted in the wave amplitude $A_0$ than the KP limit. Tanaka experienced that it took long propagation to achieve the saturated conditions in stem amplitude ($X\sim100-300$). Like other laboratory or numerical experiments (e.g. \cite{F:80},\cite{KTK:98},\cite{BK:99},\cite{P:57},\cite{M:80}), his model fails to simulate the four-fold amplification of the stem wave. The maximum stem-wave amplification computed by Tanaka \cite{T:93} is $\alpha = 2.90$ at $k = 1.03$ in the case of $A_0 = 0.28$.

Note that Tanaka's maximum amplification $\alpha = 2.90$ is presented at $k = 0.695$  instead of 1.03. It is because Tanaka followed Miles's analysis \cite{M:77b}, which assumes the incident wave angle $\psi_0$ to be small, replacing $\tan\psi_0\rightarrow\psi_0$. Tanaka did not make the adjustment for the phase (see (\ref{Heta}) on Section 2), and in his paper, $k$ is defined as $k = \psi_0/\sqrt{3a_i}$, where $a_i$ is the ratio of the incident wave amplitude $a_0$ to the reference flow depth $h_0$. Tanaka's maximum amplification occurs when $\psi_0 = 37.8^{\circ}$ and $a_i$ = 0.3. Evidently the incident wave angle is too large to make the approximation of $\tan\psi_0 \approx \psi_0$. Furthermore, the phase adjustment made for the KP limit in (\ref{Heta}) cannot be considered negligible. To reflect the finite values of incident angle and to ensure the incident wave to be a KdV line-soliton, one must use $k=\tan\psi_0/\sqrt{2A_0}$ where $A_0$ is defined by \eqref{amp}, that is, 
the $k$-parameter should be
\begin{equation}\label{newk}
 k=\frac{\tan\psi_0}{\sqrt{3(\hat{a}_0/h_0)}\cos\psi_0}.
 \end{equation}
  As we discussed in Section 2, the wave amplitude $\hat{a}_0$ in this way represents that of the corresponding incident line-soliton, i.e. $a_i=\hat{a}_0/h_0$. The difference in the value of $k$ is significant when the value of $\psi_0$ is not small, obviously.

\begin{figure}[h]
\centering
\subfloat[][]{\includegraphics[width=2.3in,height=1.9in]{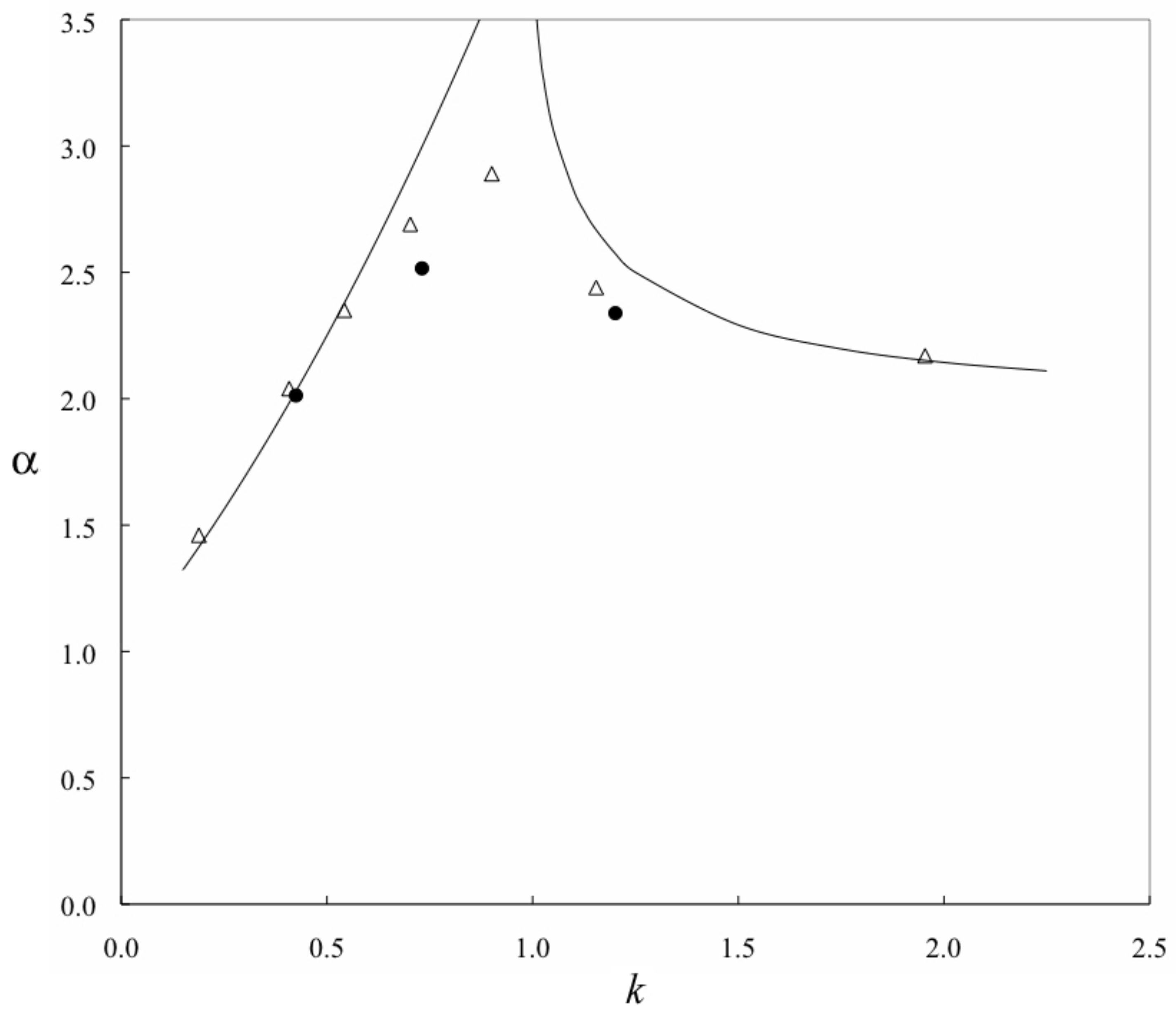}\label{fig:Fig10a}}
\hskip 0.5cm
\subfloat[][]{\includegraphics[width=2.3in,height=1.9in]{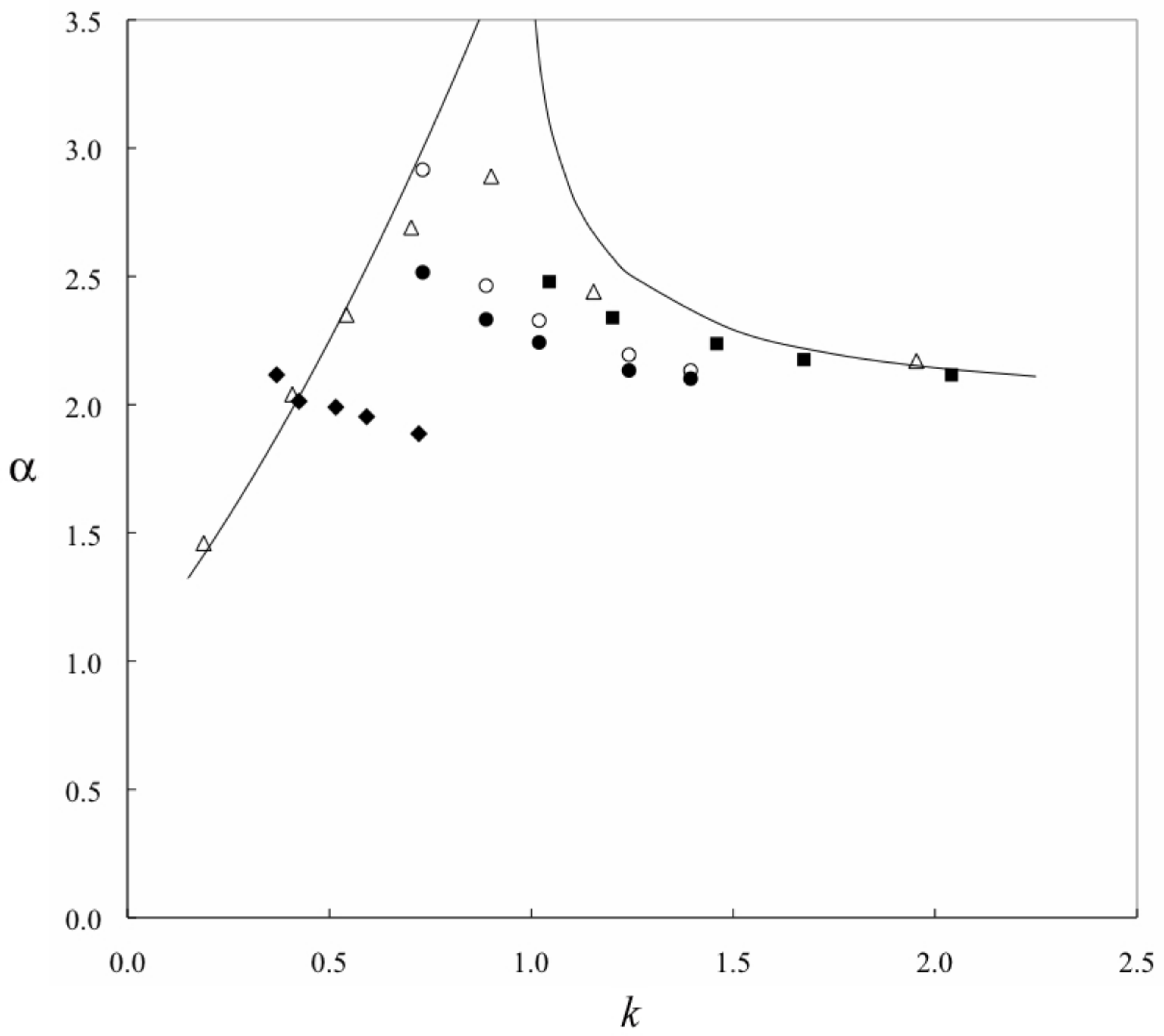}\label{fig:Fig10b}}
\caption{Comparison of the present laboratory data of the stem-wave amplification with the previous numerical results given by Tanaka \cite{T:93}, and ---, theoretical prediction of (\ref{Omax}) and (\ref{stemA}): (a) the laboratory data $\bullet$ at $X=71.1$ with the similar conditions ($\psi_0$ and $A_0$) as the numerical data at $X=150$ marked by $\triangle$; (b) the present data are at $X = 71.1$ and the incident wave angle $\psi_0= 40^{\circ}$, $\blacksquare$; $\psi_0= 30^{\circ}$, $\bullet$; $\psi_0=20^{\circ}$, $\blacklozenge$; $\psi_0=30^{\circ}$ at $X \to +\infty$, $\circ$.
The $\triangle$'s are the numerical results as in (a).} \label{fig:Fig10}
\end{figure}

The comparison with the numerical results given by Tanaka \cite{T:93} is shown in Figure \ref{fig:Fig10a}. Note that his numerical experiments were performed for a wide range of the parameter $k$ by changing the angle of incidence $\psi_0$ for a finite amplitude wave with $a_i = 0.3$. His numerical experiments fails to achieve the critical amplification of 4 at $k = 1.0$. Also plotted in Figure \ref{fig:Fig10a} are the present laboratory data of which parameters are comparable with Tanaka's data: we select the data taken at the farthest location $X = 71.1$ with the incident wave amplitude $a_i = 0.277$ and incident wave angle $\psi_0 = 20^{\circ}, 30^{\circ}$, and $40^{\circ}$ (or equivalently, $A_0$ = 0.367, 0.312, and 0.244, respectively). The laboratory data are in good agreement with Tanaka's numerical results. It is noted that the observed stem-wave amplitudes are still growing at $X$ = 71.1 as shown in Figure \ref{fig:Fig8}. 

Unlike the limited spatial domain constrained by the laboratory wave tank, numerical simulations do not have as severe limitations in space and time as the laboratory environment: Tanaka's  data \cite{T:93} are almost asymptotically stable at $X = 150$. Although there are slight differences in values of $A_0$ and $X$, Figure \ref{fig:Fig10a} demonstrates that the present laboratory results are in good agreement with Tanaka's numerical results: amplification of the numerical predictions are slightly greater than the laboratory data, because of the greater incident wave amplitude and the asymptotic amplification in Tanaka's numerical results. Most importantly, Tanaka's numerical results appear to be in good agreement with \eqref{Omax} and \eqref{stemA} of the KP theory as shown in Figure \ref{fig:Fig10a}, except near the critical point $k = 1.0$, in spite of the large incident wave angles $\psi_0$. It is emphasized that our interaction parameter $k$ based on the KP theory ($k=\tan\psi_0/\sqrt{2A_0}$) is the one that leads to significant improvement in the comparison. As discussed earlier, Tanaka's  numerical data were significantly deviated to the lower values of $k$, because
instead of \eqref{newk}, he used Miles's  parameter $k=\psi_0/\sqrt{3a_i}$ defined in \cite{M:77a},\cite{M:77b}. While $\tan\psi_0/\sqrt{2A_0}\approx\psi_0/\sqrt{3a_i}$ for infinitesimal $\psi_0$, it is critical to make proper interpretations for the approximated parameters when the mathematical theory is compared with the experimental data with finite values of $\psi_0$. None the less, the good agreement shown in Figure \ref{fig:Fig10a} is surprising for the large $k$ with large incident wave angles. One should emphasize that those results for large $k$ values  fit better in the KP theory, but not in the theory of regular
non-grazing reflection described in \cite{F:80},\cite{T:93}.  In fact, Funakoshi's results \cite{F:80} with a small
amplitude incident wave are in excellent
agreement with the KP theory for all $k$ values with this new interpretation of the $k$-parameter.

Figure \ref{fig:Fig10b} shows the stem wave amplification measured in the laboratory at $X = 71.1$ (our farthest measuring location) with $\psi_0 = 20^{\circ}$ and $30^{\circ}$, and at $X = 61.0$ with $\psi_0 = 40^{\circ}$; also plotted are the KP prediction (\ref{Omax}), (\ref{stemA}), and Tanaka's numerical results. Note that in the case of $\psi_0 = 40^{\circ}$, the data at $X = 71.1$ are slightly contaminated due to the flow contraction owing to the limited breadth of the wave tank; hence the data at $X = 61.0$ are used instead. Also plotted in the figure are the estimated asymptotic amplifications that are extrapolated from the laboratory data as shown in Table \ref{Table1}. The present laboratory data with the incident wave angle $\psi_0= 40^{\circ}$ show excellent agreement with \eqref{Omax} of the O-type solution ($k>1$) as shown in Figure \ref{fig:Fig10b}. However, as the incident wave angle decreases, the measured amplification deviates from the theoretical prediction. It is surprising to observe in Figure \ref{fig:Fig10b}  that for the (3142)-type ($k < 1$) the amplification predicted by (\ref{stemA}) should increase as the value of $k$ increases; our laboratory results show opposite. The laboratory data show no sign of transition from the (3142)-type to the O-type reflection. All the present laboratory results appear to exhibit the characteristics of Mach reflection, i.e. a) the reflected wave angle is larger than the incident angle; b) the stem length continually grows. The smaller the incident wave amplitude, the smaller the amplification factor $\alpha$ of the stem wave, and the farther the deviation from the theoretical prediction of (\ref{Omax}) for $k < 1$. 

\begin{figure}[t]
\begin{center}
\subfloat[]{\includegraphics[width=2.2in,height=2.2in]{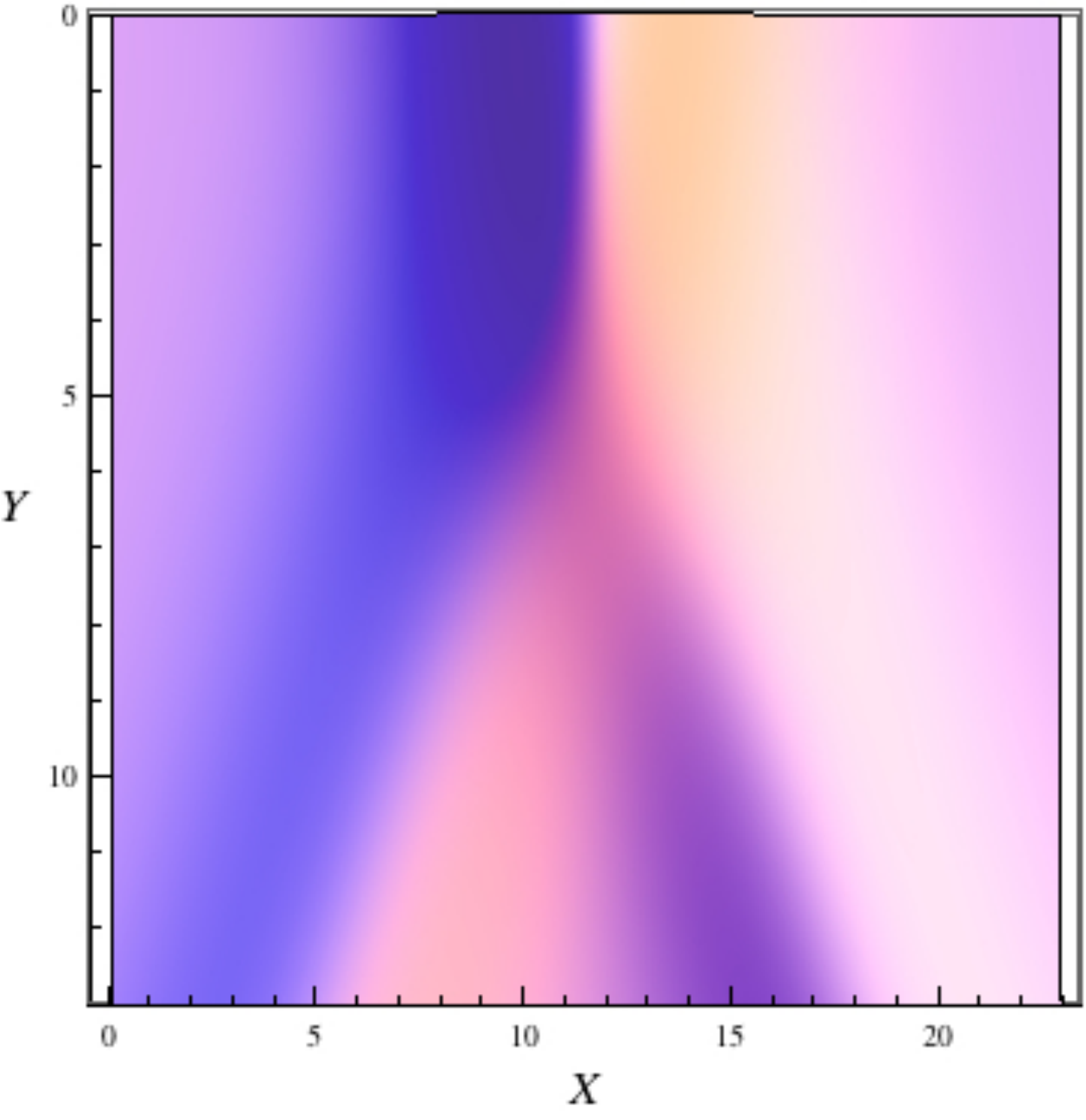} \hskip 0.8cm \includegraphics[width=2.2in,height=2.2in]{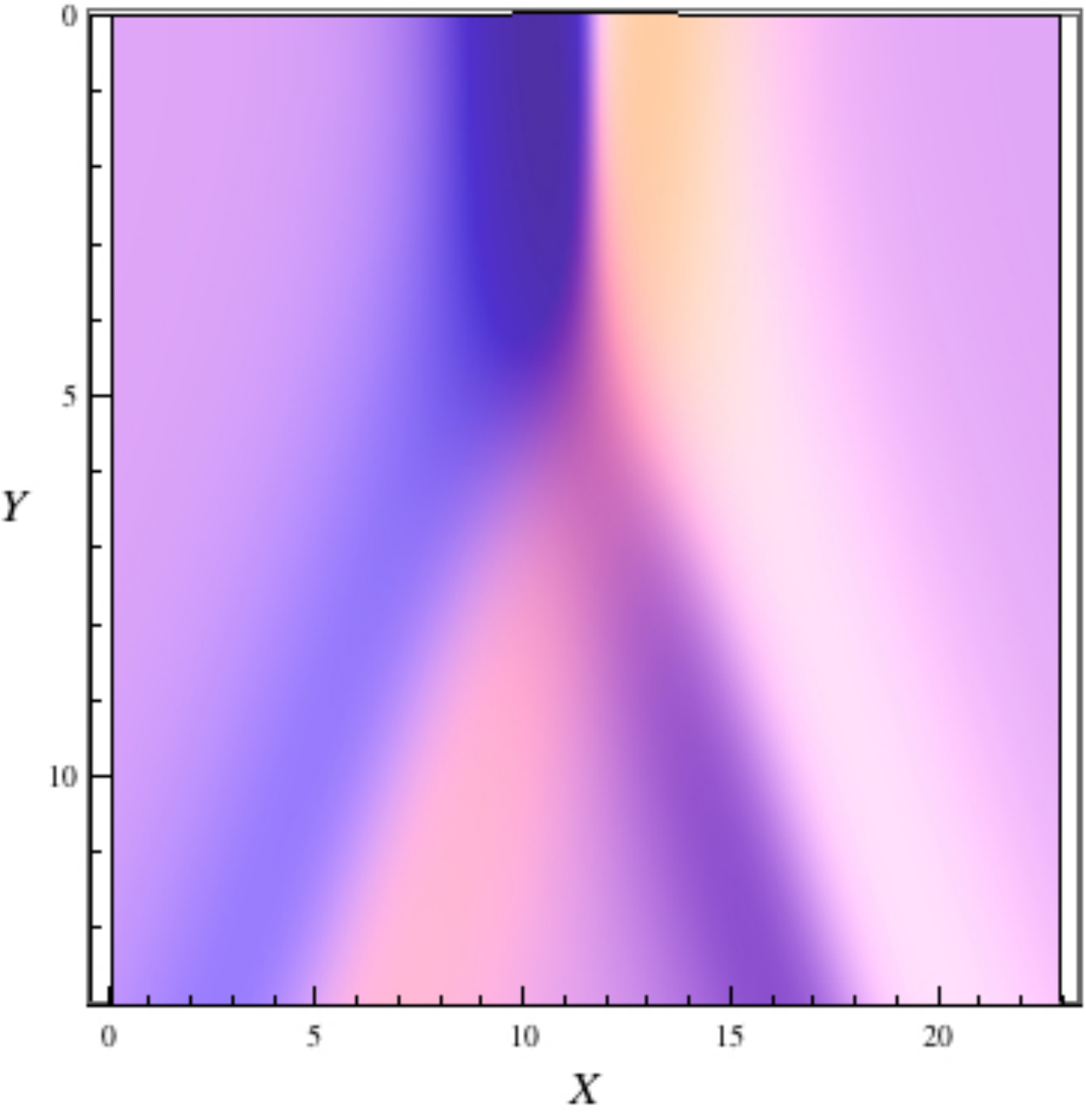}
\label{fig:Fig11a}}\\
\subfloat[]{\includegraphics[width=2.2in,height=2.2in]{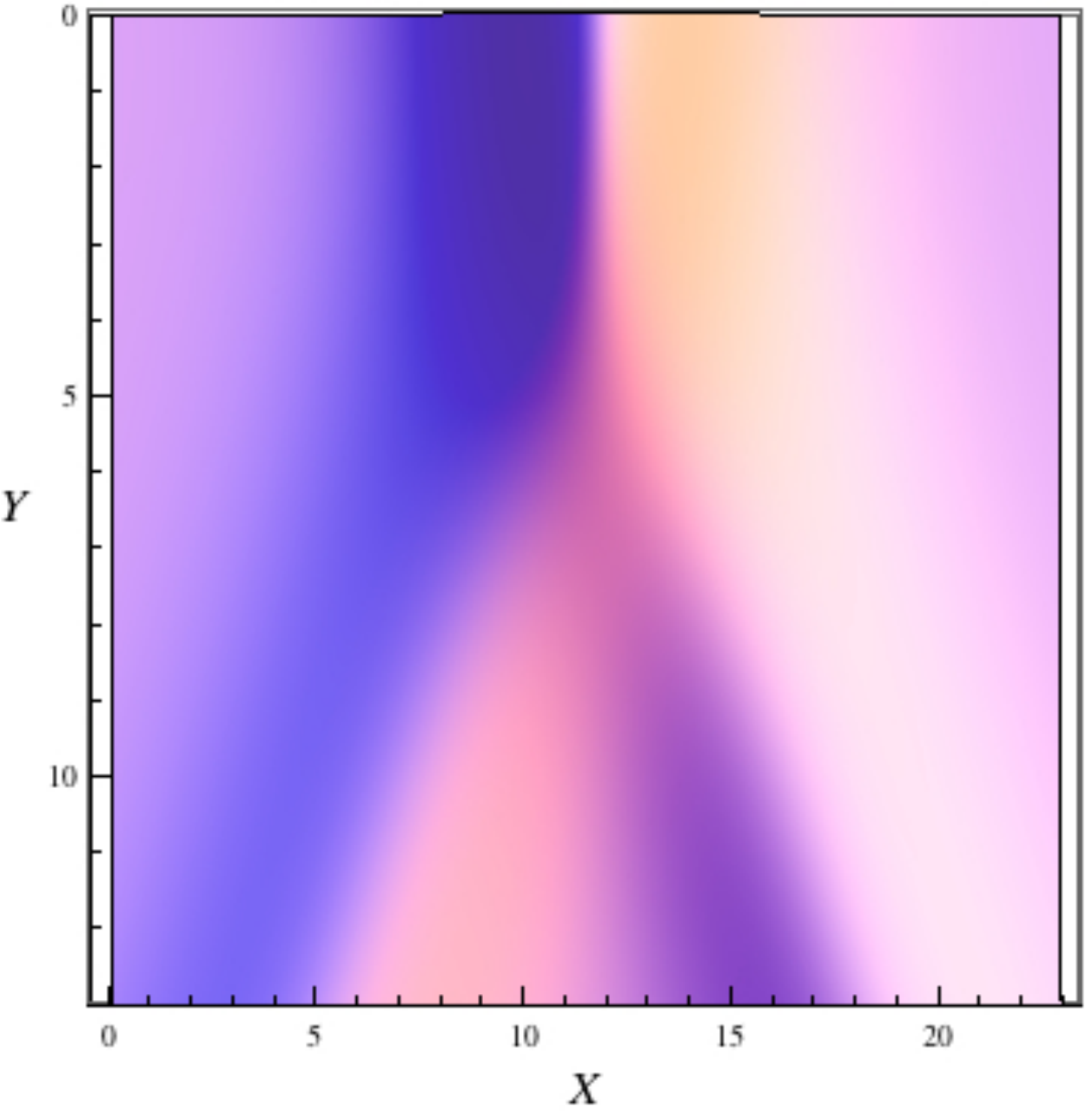} \hskip 0.8cm \includegraphics[width=2.2in,height=2.2in]{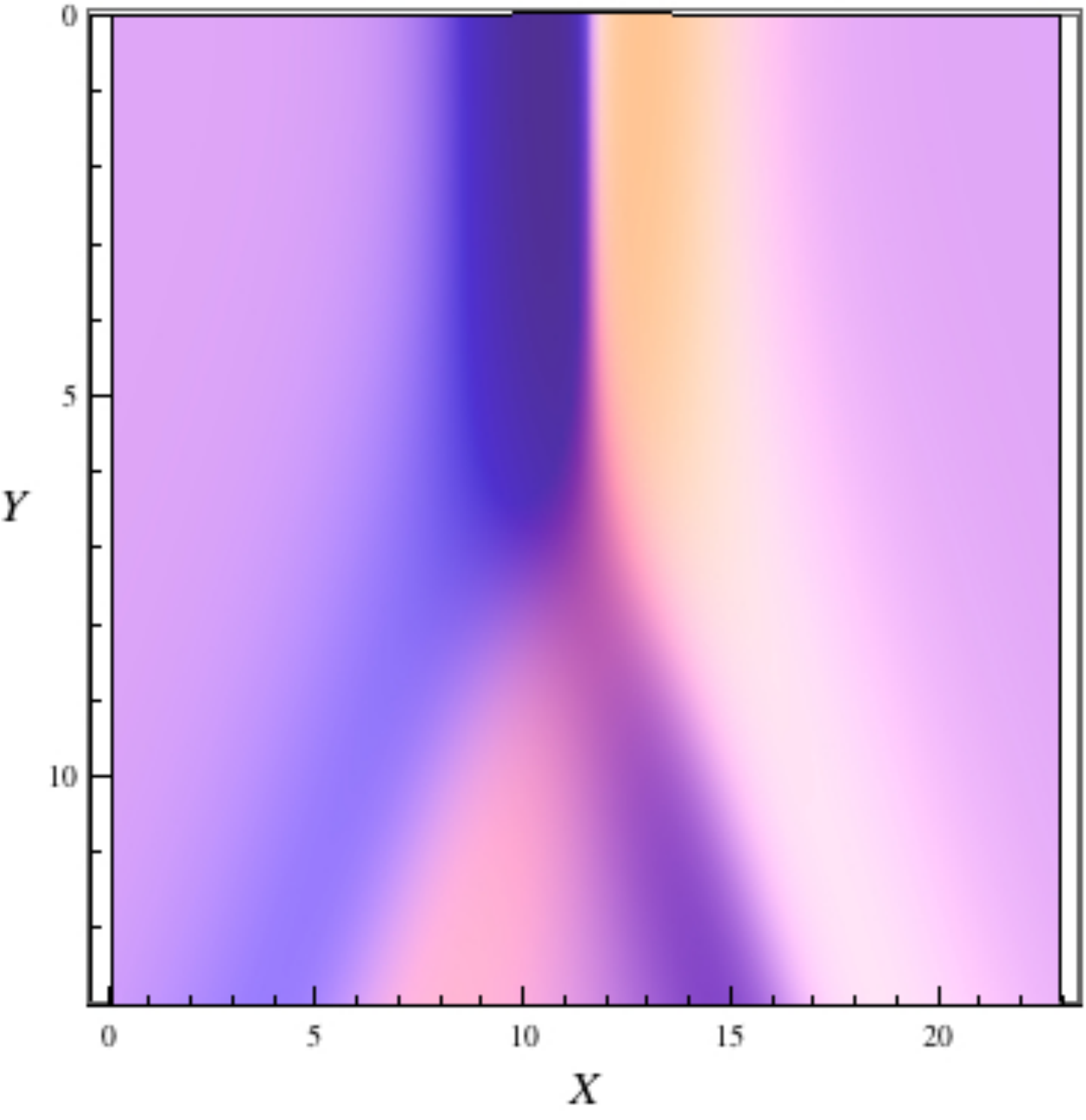}
\label{fig:Fig11b}}
\caption{Solutions of the KP equation for $\psi_0= 30^{\circ}$: (a) at $X = 71.1~ (T=41.05)$, $A_0$ = 0.112, the O-type (left) and $A_0$ = 0.212, the (3142)-type (right);  (b) at $X = 208~ (T=120)$, $A_0$ = 0.112, the O-type (left) and $A_0$ = 0.212, the (3142)-type (right).
Notice that the stem parts of both those solutions are similar at $X=71.1$, but
later the $(3142)$-solution has longer stem, the Mach stem.}\label{fig:Fig11}
\end{center}
\end{figure}

A possible explanation for the large discrepancy that appears in the case of $\psi_0 = 20^{\circ}$ may be due to the limitation of propagation distance in the laboratory. With a small incidence angle, the development of Mach reflection may take a very long distance to reach its asymptotic state. Examination of our KP solutions supports this. Figure \ref{fig:Fig11a} shows the KP solutions at $X = 71.1$ based on the recipe provided in Section 2 for the cases with $A_0$ = 0.112 and 0.212 with $\psi_0 = 30^{\circ}$. Note that the former case $A_0$ = 0.112 is the O-type because $\psi_0 > \psi_c = 25.3^{\circ}$, and the latter case $A_0$ = 0.212 is the (3142)-type because $\psi_0 < \psi_c = 33.1^{\circ}$. The same waves but at $X = 208$ are presented in Figure \ref{fig:Fig11b}. At $X = 71.1$, the stem length of the O-type solution appears similar to that of the (3142)-type, while they are distinctively different at $X = 208$. This observation indicates that both the O-type and the (3142)-type would grow their stems at the similar rate at least up to $X = 71.1$. However, the growth of the stem is ceased for the O-type once it reaches its equilibrium state. Evidently, this is not the Mach stem formation described by Miles \cite{M:77b}. On the other hand, the Mach stem of the (3142)-type appears continual growth (see Figure \ref{fig:Fig11} right). Most of our laboratory data were recorded at $X = 71.1$, where both O-type and (3142)-type are still being developed. The KP solutions in Figure \ref{fig:Fig11} suggest that the laboratory tank must be too short to distinguish these two types of the KP solutions.

Table \ref{Table2} shows the KP predictions of amplification $\alpha$ with $\psi_0 = 20^{\circ}$ at $X = 71.1$; the predicted amplifications are in excellent agreement with the measured values. We extended the KP predictions to a longer distance, $X = 350$, close to the asymptotic amplifications. The excellent agreement in amplification of the predictions with the measurements at $X = 71.1$ suggests that the shortcoming of the laboratory experiments should be attributed to the available propagation distance that must be insufficient to establish its asymptotic condition. 

\begin{table}[th]
\centering
\caption{Asymptotic amplification of the stem waves with $\psi_0=20^{\circ}$:  
$k=\tan\psi_0/\sqrt{2A_0}$, and 
$\alpha_{X = 71.1}(\rm Exp.)$ is the measured stem-wave amplification.
$\alpha_{X = 71.1} (\rm KP)$ is the KP prediction at $X$ = 71.1, and $\alpha_{X = 350} (\rm KP)$ is that at $X$ = 350 where the reflection should become its asymptotic state. \label{Table2}}
\begin{tabular}{ccccc}
\hline
\multicolumn{1}{c}{$A_0$} &\multicolumn{1}{c}{$k$}& \multicolumn{1}{c}{$\alpha_{X = 71.1}(\rm Exp.)$} & \multicolumn{1}{c}{$\alpha_{X = 71.1} (\rm KP)$} & \multicolumn{1}{c}{$\alpha_{X = 350} (\rm KP)$}\\
\hline
0.127 & 0.722& 1.89 & 1.84 & 2.87\\
0.189 & 0.591& 1.95 & 1.93 & 2.52\\
0.249 & 0.516& 1.99 & 2.08 & 2.33\\
0.367 & 0.425& 2.01 & 1.99  &2.03\\
\hline
\end{tabular}
\end{table}

\section{Summary and Conclusions}\label{sec:4}
 
Amplification of obliquely incident solitary wave onto a vertical wall was analyzed based on the KP equation. The KP equation is equivalent to the approximation of the theory developed by Miles \cite{M:77a},\cite{M:77b}. Consequently the same predictions result for the asymptotic conditions, which include the maximum four-fold amplification at the critical state. Unlike Miles's theory, however, the KP equation enables us to analyze the development processes. Our theoretical predictions were analyzed with our laboratory experiments as well as the previous numerical experiments by Tanaka \cite{T:93}.

Careful attention must be taken when theoretical predictions are compared with laboratory or numerical data. Adequate adjustments must be made to link the physical quantities with the mathematical parameters. Specifically, the interaction parameter $k$ should be computed with 
\eqref{newk}, i.e. $k=\tan\psi_0/\sqrt{2A_0}=\tan\psi_0/(\sqrt{3a_i}\cos\psi_0)$, instead of $k = \psi_0/\sqrt{3a_i}$. The former parameter $k$ ensures that the incident wave is a KdV soliton, while the later would not. It is because the incident wave angle $\psi_0$ is not infinitesimal but finite in the experiments. Discrepancy in amplification reported in Tanaka's paper \cite{T:93} improves significantly with the use of proper $k$ as demonstrated in Figure \ref{fig:Fig10a}. We also confirm that Funakoshi's results \cite{F:80} with smaller incident amplitudes are in an excellent agreement with the KP prediction with our interaction parameter $k$.

The present laboratory results of wave amplification are consistent with Tanaka's numerical results if and only if the data with similar conditions are compared. However, when the comparison is extended to include all other data, the present results turn out different from both KP theory and Tanaka's numerical results. The KP theory indicates that our laboratory apparatus is too short to demonstrate the difference between the O-type and the (3142)-type solutions. In fact, the stem wave along the wall grows at the similar rate for both O-type and (3142)-type along the available length in the laboratory. The theory predicts that the growth rate soon deviates beyond the distance of the apparatus -- the O-type solution ceases the growth when it reaches its equilibrium state, while the stem wave continually grows for the (3142)-type as the characteristic of the Mach stem. Considering a long distance (time) for an obliquely incident soliton to become the asymptotic state, especially for a small incidence $\psi_0$, it would be difficult to realize the theoretical predictions in a real-fluid laboratory environment.

\begin{acknowledgement}
This work is supported by Oregon Sea Grant Program No. NA06OAR4170010.
One of the authors (Y.K.) is partially supported by NSF Grant DMS-0806219.
\end{acknowledgement}

\end{document}